\documentstyle[12pt,epsf]{article} 
\newcommand{\vp}{\varphi}

\newcommand{\HH}{{\cal H}}
\newcommand{\MM}{{\cal M}}

\newcommand{\TT}{{\cal T}}
\newcommand{\VV}{{\cal V}}

\newcommand{\wh}{\widehat}

\newcommand{\be}{\begin{equation}}
\newcommand{\ee}{\end{equation}}
\newcommand{\ben}{\begin{eqnarray}\displaystyle}
\newcommand{\een}{\end{eqnarray}}
\newcommand{\refb}[1]{(\ref{#1})}

\newcommand{\sectiono}[1]{\section{#1}\setcounter{equation}{0}}
\renewcommand{\theequation}{\thesection.\arabic{equation}}

\catcode`\@=11
\def\marginnote#1{}
\newcount\hour
\newcount\minute
\newtoks\amorpm
\hour=\time\divide\hour by60
\minute=\time{\multiply\hour by60 \global\advance\minute by-\hour}
\edef\standardtime{{\ifnum\hour<12 \global\amorpm={am}%
        \else\global\amorpm={pm}\advance\hour by-12 \fi
        \ifnum\hour=0 \hour=12 \fi
        \number\hour:\ifnum\minute<10 0\fi\number\minute\the\amorpm}}
\edef\militarytime{\number\hour:\ifnum\minute<10 0\fi\number\minute}
\def\draftlabel#1{{\@bsphack\if@filesw {\let\thepage\relax
   \xdef\@gtempa{\write\@auxout{\string
      \newlabel{#1}{{\@currentlabel}{\thepage}}}}}\@gtempa
   \if@nobreak \ifvmode\nobreak\fi\fi\fi\@esphack}
        \gdef\@eqnlabel{#1}}
\def\@eqnlabel{}
\def\@vacuum{}
\def\draftmarginnote#1{\marginpar{\raggedright\scriptsize\tt#1}}
\def\draft{\oddsidemargin -.5truein
        \def\@oddfoot{\sl preliminary draft \hfil
        \rm\thepage\hfil\sl\today\quad\militarytime}
        \let\@evenfoot\@oddfoot \overfullrule 3pt
        \let\label=\draftlabel
        \let\marginnote=\draftmarginnote
   \def\@eqnnum{(\theequation)\rlap{\kern\marginparsep\tt\@eqnlabel}%
\global\let\@eqnlabel\@vacuum}  }

\def\preprint{\twocolumn\sloppy\flushbottom\parindent 1em
        \leftmargini 2em\leftmarginv .5em\leftmarginvi .5em
        \oddsidemargin -.5in    \evensidemargin -.5in
        \columnsep 15mm \footheight 0pt
        \textwidth 250mmin      \topmargin  -.4in
        \headheight 12pt \topskip .4in
        \textheight 175mm
        \footskip 0pt
        \def\@oddhead{\thepage\hfil\addtocounter{page}{1}\thepage}
        \let\@evenhead\@oddhead \def\@oddfoot{} \def\@evenfoot{} }

\def\titlepage{\@restonecolfalse\if@twocolumn\@restonecoltrue\onecolumn
     \else \newpage \fi \thispagestyle{empty}\c@page\z@
        \def\thefootnote{\fnsymbol{footnote}} }

\def\endtitlepage{\if@restonecol\twocolumn \else  \fi
        \def\thefootnote{\arabic{footnote}}
        \setcounter{footnote}{0}}  %\c@footnote\z@ }

\catcode`@=12
\relax
\def\bea{\begin{array}}
\def\bem{\begin{displaymath}}
\def\beq{\begin{equation}}

\def\bra#1{\left\langle #1\right|}
\def\eea{\end{array}}
\def\eem{\end{displaymath}}
\def\eeq{\end{equation}}
\def\half{\frac{1}{2}}

\def\ket#1{\left| #1\right\rangle}

\def\s2w{\sin^2 \theta_W}

\newcommand{\real} {{{\rm I} \kern -0.2em {\rm R}}}
\newcommand{\complex} {{{\sf I} \kern -0.48em {\rm C}}}
\newcommand{\naturel} {{{\rm I}  \kern -0.18em {\rm N}}}
\newcommand{\integer} {{{\rm Z} \kern -0.31em {\rm  Z}}}
\newcommand{\smallinteger} {{{\rm Z} \kern -0.25em {\rm  Z}}}

\relax
\def\crbig{\\\noalign{\vspace {3mm}}}

\relax
\begin{document}
%\topmargin-2.4cm
%\draft
%\preprint
\setcounter{footnote}{0}
\setcounter{page}{1}

{}~
\hfill\vbox{\hbox{hep-th/0005036}\hbox{MIT-CTP-2975}
 \hbox{MRI-PHY/P20000430}
}\break

\vskip 2.0cm

\centerline{\large \bf D-branes as  Tachyon
Lumps in String Field Theory}

\vspace*{6.0ex}

\centerline{\large \rm Nicolas Moeller$^a$, Ashoke Sen$^b$ and Barton
Zwiebach$^c$}

\vspace*{6.5ex}

\centerline{\large \it $^{a,c}$Center for Theoretical Physics}

\centerline{\large \it
Massachusetts Institute of Technology}

\centerline{\large \it Cambridge,
MA 02139, USA}
\vspace*{1ex}
\centerline{E-mail: moeller@pierre.mit.edu, zwiebach@mitlns.mit.edu}

\vspace*{1ex}
\centerline{\large \it ~$^b$Mehta Research Institute of Mathematics}

\centerline{\large\it
and Mathematical Physics, Chhatnag Road,}

\centerline{\large \it   Jhoosi,
Allahabad 211019, INDIA}
\vspace*{1ex}
\centerline{E-mail: asen@thwgs.cern.ch, sen@mri.ernet.in}

\vspace*{4.5ex}

\centerline {\bf Abstract}
\bigskip
It has been conjectured that the tachyonic lump solution of the open
bosonic
string field theory
describing a D-brane
represents a D-brane of one lower dimension.
We place the lump on a circle of finite radius and
develop a variant of the level
expansion scheme that allows systematic account of all higher derivative
terms in the string field theory action,
and gives a calculational scheme that can be carried to arbitrary
accuracy.
Using this approach we
obtain lump masses
that agree with expected D-brane masses to an accuracy of about
1\%.
We find convincing
evidence that in string field theory the lump representing
a D-brane is an extended object with
a definite profile. A gaussian fit to the lump 
gives a 6-sigma size of $9.3
\sqrt{\alpha'}$. The level truncation scheme developed here naturally
gives
rise to
an infrared and ultraviolet cut-off, and may be useful in the study of
quantum string field theory.

\vfill \eject

\baselineskip=18pt

\tableofcontents

\sectiono{Introduction and Summary} \label{s0}

The 26-dimensional critical bosonic string theory admits Dirichlet
$p$-branes (D-$p$-branes) for all $p\le 26$. Each of these D-$p$-branes
admits a tachyonic mode $T$ of mass$^2=-1$, in units where the tension of
the fundamental string is equal to $(2\pi)^{-1}$ ($\alpha'=1$). It has
been conjectured that the potential for the tachyon field has a
non-trivial translationally invariant (local) minimum at some value
$T_{vac}$
where the sum of the tachyon potential and the tension of the original
brane vanishes~\cite{9902105}. Thus at $T=T_{vac}$ the total energy
density
vanishes, and hence this configuration can be identified as the vacuum of
the closed string theory without any D-branes.  It has also been
conjectured that although this vacuum does not have any perturbative open
string excitations, it contains lump-like soliton configurations which
approach the vacuum $T=T_{vac}$ asymptotically far away from the core of
the
soliton and represent D-branes of lower dimension~\cite{RECK,9902105}.  
Similar conjectures have also been made involving the tachyon living on
the coincident D-brane anti-D-brane pair, or on a non-BPS D-brane of type
IIA
and IIB superstring
theories~\cite{9805019,9805170,9808141,9810188,9812031,9812135}.

Various pieces of evidence for these conjectures have been found in both
the first~\cite{RECK,9902105,9805019,9805170,9808141}, and
second
\cite{9912249,0001201,0002237,0002117,0003031,0001084,0002211,0003220,0004015}
quantized
string theory, and also using AdS/CFT
correspondence~\cite{0001143,0004131}. The first
quantized description has been successful in
verifying the conjectures relating the tachyonic solitons to lower
dimensional D-branes, but it can only supply indirect evidence for the
equivalence between the (local) minimum of the tachyon potential and the
vacuum without a D-brane. On the other hand, the second quantized
description $-$ open string field theory~\cite{WITTENBSFT} $-$ can
provide direct evidence for this conjecture by explicitly computing the
(negative) value of the tachyon potential at the minimum and comparing it
with the tension of the original D-brane system. Although open string
field theory contains infinite number of fields, and the problem of
finding a translationally invariant stationary point of the potential
involves solving the equations of motion of the infinite number of zero
momentum modes of these fields, the calculations are made feasible by
using the level expansion scheme proposed by Kostelecky and
Samuel~\cite{KS}.  The procedure is as follows.  Using the correspondence
between the modes of the string field and states in the conformal field
theory describing the first quantized string, we define the level of a
mode of the string field as the difference between the $\wh N$
eigenvalue
of the first quantized string state representing this mode, and the $\wh
N$
eigenvalue of the state representing the zero momentum
tachyon, where $\wh N$ is the total `number operator' of the
matter and ghost system.
The level truncation scheme to order $(M,N)$
then corresponds to an approximation in which we keep in the string field
theory action all modes of level $\le M$, and all interaction terms for
which the sum of the levels of all the modes appearing in the term is $\le
N$. This gives a potential (which, for a static field configuration, is
just the negative of the action up to a normalization constant) with
finite number of fields and a finite number of terms. Thus we can find its
extremum and calculate its value at the extremum. The larger the values of
$(M,N)$, the larger is the number of modes and the number of terms in
the
potential, and the better is the accuracy.

The calculation of ref.~\cite{KS} for the tachyon potential was revisited
and extended in~\cite{9912249} in terms of background independent fields.
It was shown there that the total negative potential energy at the
stationary point cancels the energy of the D-brane represented by the
string field theory to an accuracy of $<$1.5\% at the level (4,8)
approximation. This calculation was extended in ref.~\cite{0002237} to
level (10,20). At this level the
contribution from the tachyon potential was found to cancel the tension of
the D-brane to an accuracy of about .1\%.  Similar calculations have also
been
performed~\cite{0001084,0002211,0003220,0004015} in open
superstring field theory~\cite{9503099,9912121,9912120}. At the level
(2,4)
approximation the tachyon potential has been shown to cancel about 90\% of
the
tension of the original brane configuration.\footnote{At present there
seems to be some disagreement between refs.~\cite{0003220} and
\cite{0004015} about the level (2,4) results.}

The success of string field theory in verifying the conjecture relating
the translationally invariant stationary point of the tachyon potential
and the vacuum without any D-brane encourages one to ask whether string
field theory can also be used in studying the conjectured relation between
the tachyonic lump solutions and lower dimensional D-branes. This study
was initiated by Harvey and Kraus~\cite{0002117}. In this paper they
started with the level (0,0) contribution to the tachyon potential in open
bosonic string field theory on a D-$p$-brane, and identified a `bounce
solution' in this field theory as the D-$(p-1)$ brane. At this level the
tension associated with this solution turns out to be about 78\% of the
known value of the D-$(p-1)$-brane tension.  This result receives
correction not only from the higher level fields, but also from the
momentum dependence of the interaction terms which were neglected in the
initial analysis. While there is no systematic expansion scheme
for taking into account these momentum dependent corrections, a naive
expansion of the interaction term in powers of momentum, keeping only
the zeroth and first order terms, reduced the tension of
the soliton to about 70\% of the conjectured answer. On the other hand,
taking into account the correction to the potential to level (2,4)
increased the answer back to about 82\% of the conjectured answer. A
systematic
method for taking into account the momentum dependent terms in the
interaction was suggested in ref.~\cite{0003031}, but this procedure did
not give rise to an appreciable change in the tension of the lump.
A similar
analysis has also been carried out for solitons in the open superstring
field theory~\cite{0002211,berg,amer}. Although the answer turns out to
be close
to the expected answer, it is likely to be an accidental result, as there
is no reason to assume that the corrections due to the momentum dependent
terms are small in this case.

The purpose of this paper will be to develop a systematic approximation
scheme for studying these solitons in string field theory and calculating
their tension. We shall focus on the codimension one lump on a D$p$-brane
of the bosonic string theory $-$ which is conjectured to be equivalent to
a D-$(p-1)$-brane $-$ but it will become clear that the scheme is general
enough to be applicable to the study of higher codimension solitons, as
well as to solitons in superstring field theory. In the case of
a codimension one soliton, we are dealing with a field configuration on
the
D$p$ brane which depends on only one of the spatial coordinates (say $x$)
on the brane, and is independent of time, as well as the other $(p-1)$
spatial coordinates. We study this problem by compactifying the coordinate
$x$ on a circle of radius $R$ instead of letting it span the whole real
line. In this case, since all field configurations must be periodic in
$x$, we can decompose all fields into modes carrying discrete momenta
along $x$ in units of $(1/R)$, and the solitonic field configuration that
we are looking for must be obtained as an appropriate superposition of
these modes. We can now define the level of any such mode as the
difference between the $L_0$ eigenvalue of the first quantized string
state representing this mode, and that of the zero momentum tachyon 
state, where $L_0$ denotes the zeroth component of the
Virasoro generator of the combined matter ghost system.\footnote{Since for
the zero momentum states the eigenvalue of the number operator is the
same as the $L_0$ eigenvalue of the state, the two prescriptions agree
for these states.} 
This allows us to define a level $(M,N)$ approximation to the potential
exactly as before.  Working with the potential up to a given level, we can
now look for $x$ dependent solutions of the string field equations by
extremizing the potential with respect to the modes appearing in the
potential to this level.

This is precisely the procedure we follow in this paper for studying the
tachyonic lump solution on a D-$p$-brane.\footnote{The discretization of
the momentum is reminiscent of the procedure followed in
ref.~\cite{0003031}, although the precise relationship between these two
approaches is not clear.} We study this problem for various radii at
various levels of approximation, and compare the tension of the lump with
the tension of a D-$(p-1)$-brane.  The results for the tension of the lump
turn out to be remarkably close to the known tension of the
D-$(p-1)$-brane. Whereas for $R=\sqrt{3}$ and $\sqrt{15/2}$ we are able to
get a lump tension within 1\% of the tension of the D-$(p-1)$-brane, for
larger radii ($R=\sqrt{12}$ and $\sqrt{35/2}$) we get answers within 3\% of
the expected answer. We also compare the profile of the tachyon field
corresponding to the lump for different values of $R$, $-$ obtained by
superposition of $\cos(n x / R)$ for integer $n$ $-$ and find remarkable
agreement between the profiles for different values of $R$.

At this point we should note that the problem of formation of the
tachyonic lump on a circle was addressed using the first quantized
approach in ref.~\cite{0003101}. There a
renormalization group analysis was used to show that
the
mass of the
tachyonic lump on a D-$p$-brane is indeed equal to that of a
D-$(p-1)$-brane.\footnote{This followed earlier work
of ref.~\cite{9406125} on the renormalization group flow of the two
dimensional field theory under a
perturbation corresponding to switching on a
tachyon background proportional to $\cos(x/R)$.}  In view of this result
one might ask whether
the string
field theory analysis carried out in this paper gives any new insight into
this problem. To this end, we note, first of all, that the relationship
between the renormalization group analysis in the first quantized
approach, and the string field theory analysis based on the level
truncation scheme, is as yet quite unclear, and hence it is certainly
illuminating to independently verify the equivalence of the
D-$(p-1)$-brane, and the tachyonic lump on the D-$p$-brane in string field
theory. Furthermore, string field theory provides us with a definite
picture of the tachyon profile as superposition of $\cos(n x / R)$ for
different $n$ with definite coefficients. In contrast the analysis based
on the renormalization group flow only tells us that a perturbation by the
leading relevant operator $\cos(x/R)$ takes the original D-$p$-brane to a
D-$(p-1)$-brane, and does not tell us how the higher harmonics mix with
$\cos(x/R)$ to produce the soliton. Indeed most of the higher harmonics
correspond to irrelevant perturbation, and hence their coefficients vanish
in the infra-red.\footnote{Presumably if we could determine the exact
location of the infrared fixed point in the space of coupling constants,
then the shape of the lump will be determined in this
approach.}
Furthermore, the rigorous results of ref.\cite{9406125} have not yet been
generalized to superstring theory.
Thus we believe that despite the exact
results based on the renormalization group analysis of the first quantized
theory, the present analysis throws new light on the tachyonic soliton
solutions.

The rest of the paper is organized as follows. In section~\ref{s1} we
outline the general procedure of level expansion scheme of the string
field theory, discuss the possibility of restricting the string field
to a background independent subspace for studying the lump solution, and
give details of computation of a few terms in
the potential. In section~\ref{s2} we give in detail the results for the
potential, the lump
solution and its energy for a specific radius $R=\sqrt{3}$. We also
compare the profile of the lump at different levels of approximation. In
section~\ref{s3} we give the results for several other radii, both larger
($\sqrt{15/2}$, $\sqrt{12}$ and $\sqrt{35/2}$) and smaller
($\sqrt{11/10}$) than
$\sqrt{3}$, and compare the profile of the lump for each radii with the
profile at $R=\sqrt{3}$. We conclude in section~\ref{s4} by discussing
possible generalization of this analysis and some speculations.

\sectiono{Level Expansion and the String Field} \label{s1}

In this section we will set up a variant of the level
expansion method to deal with the problem of finding
the profile and mass of the tachyon lump in string field
theory. As reviewed in the introduction, such method
is desirable as previous computations of lump masses
in string field theory have not been very accurate.  After
explaining this method we will discuss the background
independent expansion of the string field suitable for
the problem.  Then we discuss two methods for estimating
the lump mass.  We conclude by showing a few samples
of typical calculations needed to evaluate the string field
action for the lump.

\subsection{Modified level expansion}

When calculating the tachyon potential in search for
a spacetime independent vacuum state, all spacetime
fields are set to constants, and the evaluation of 
the string field action does not require the inclusion
of terms with spacetime derivatives. The string field
is at zero momentum and is thus built by a superposition
of zero momentum states times constants representing the 
zero momentum modes of the spacetime fields.  The states are built by 
acting on a zero-momentum vacuum with oscillators of the
relevant conformal field theory (CFT). In this case the
level expansion was defined as follows~\cite{KS}. 
Let $\wh N$ be the number operator, representing the contribution to $L_0$
from
the system of matter and ghost 
oscillators. Let $N_0$ (=$-1$) denote the
eigenvalue of $\wh N$ for the zero momentum tachyon: $\wh N\ket{T_0} = N_0
\ket{T_0}$.
For a given state
$\ket{\Phi_i}$,  with number eigenvalue $N_i$ ($\wh N \ket{\Phi_i} = 
N_i \ket{\Phi_i}$)
we define the level $l(\Phi_i)$ of the state $\ket{\Phi_i}$
as
\begin{equation}
l(\Phi_i)  \equiv N_i - N_0\,. 
\end{equation}
As defined, level is a dimensionless number. For the
case of bosonic string theory the levels are all integers
while for NS superstrings they can also be half integral.
We now define the level $(M,N)$ approximation to the action as follows:
\begin{itemize}
\item We keep only those fields with level $\le M$.
\item We keep only those terms in the action for which the sum of the 
levels of
all the fields in the term is $\le N$.
\end{itemize}
In order that the quadratic term of all fields with level $\le M$ are
kept in the action, we must have $N\ge 2M$. 
While variants are possible, it seems most effective 
when calculating any physical object to use its level $(M,2M)$ 
approximation, as experience shows that increasing the number of terms in
the
potential keeping the number of fields fixed does not improve the
results very much. While there is yet no
theoretical explanation for the convergence  of the level expansion,
the numerical evidence collected thus far is impressive.

Consider now the problem at hand. While all of our discussion
applies to soliton solutions on non-BPS D-branes, and D-brane anti-
D-brane
pairs of
superstring
theory, we will consider here explicitly
only the case of the unstable D-branes of bosonic string theory. 
Consider therefore, 
an unstable bosonic
D-brane extending over a number of spatial dimensions. We now 
wish to select one of these dimensions, call it $x$ and construct a 
tachyon lump such that the
solution depends only on the $x$-coordinate. (Again our discussion applies
to lumps depending on more than one coordinates, but we shall not analyze
these cases here.) As the lump is not invariant under translation along
$x$, we now  need to include  $x$-momentum modes in 
the string field expansion
and $x$-derivatives, or $x$-momentum dependent terms in the 
string field action. In order to do this
systematically  we compactify $x$ over a circle of 
radius $R$, namely $x \sim x+ 2\pi R$.
This quantizes the $x$-momentum as $p_x= n/R$ for
integer $n$. For each of the zero momentum states $\ket{\Phi_i}$ we had
before, we now have discrete states of the type $\ket{\Phi_{i,n}}$ that 
only differ by the fact that they are built on vacua having
$x$-momentum $n/R$. For such states there is a natural generalization
of the level. This is the difference between the $L_0$ eigenvalue of the
state and that of
the zero momentum
tachyon, where $\{L_n\}$ denote the Virasoro generators of the combined
matter and ghost system. 
This is because (with $\alpha'=1$)
we have
that $L_0 = p_x^2 + \wh N$. 
For zero momentum this is 
just the previous definition.
Still denoting by $N_i$ the number eigenvalue of $\ket{\Phi_{i,n}}$ we
have 
\be \label{enumber}
l(\Phi_{i,n}) = L_0 (\Phi_{i,n}) - L_0 (T_0) = {n^2\over
R^2} + N_i - N_0\, . 
\ee
The level is still a dimensionless number as $R$ here is measured in units
of $\sqrt{\alpha'}$ (which has been set to one). 
We can now define the level $(M,N)$ approximation for the action exactly
as before. Since
the $L_0$ eigenvalue of a state plays a crucial role in the conformal
map that inserts the state into the disk representing the interaction
terms in the action, this is a natural generalization of the level
truncation scheme of ref.~\cite{KS}. This paper will present evidence that
this modified version of the level truncation scheme also works very well.

In calculating in this setup in the level $(M,2M)$ approximation
for any given radius we will have to include states $\ket{\Phi_i}\equiv
\ket{\Phi_{i,0}}$
and ``harmonics"  $\ket{\Phi_{i,n}}$, and clearly the condition
$l(\Phi_{i,n}) \leq M$ will give an upper bound on $n$ for each
$i$. This also requires 
$l(\Phi_{i,0}) \leq M$, and thus we have a finite number of modes to be
included at a given level of approximation. 
Each term in the action including modes whose sum of levels
does not exceed $2M$ is computed exactly. It is manifest that
in a cubic string field theory the level $(M,2M)$ approximation
will only require a finite number of computations\footnote{This will
also be the case for the NS superstring field theory discussed in
ref.~\cite{0001084,0002211,0003220,0004015}}.    

\subsection{Background Independent String Field}

The general setup required to study a lump
is similar to that developed in~\cite{9911116}
to study the mass of the D-brane.  To begin with, we assume that the
background space-time is the product of a (2+1) dimensional flat
space-time, labelled by a pair of space-like
coordinates $(x,y)$ and a
time like coordinate $x^0$, and an arbitrary Euclidean manifold
$\MM$ described 
by a conformal field theory of central charge 23. We take the spatial
direction $y$ to be non-compact, but $x$ to be compact with radius $R$. We
let  $X$,
$Y$ and $X^0$ denote the three scalar fields on the string world-sheet
associated with the coordinates $x$, $y$ and $x^0$. 

We now consider a D-brane with the following properties. For an
open string ending on the D-brane we put Neumann boundary condition on the
fields $X$ and $X^0$ and Dirichlet boundary condition on the  field
$Y$.\footnote{As in
ref.~\cite{9911116}, the extra non-compact direction $y$ with Dirichlet
boundary condition provides a direction along which the brane can move,
and we can calculate the tension of the brane by studying its motion in
this direction.} We
leave the boundary condition on the fields associated with the coordinates
on $\MM$ arbitrary, with the
only restriction that all the fields on which we put Neumann boundary
condition are associated with compact coordinates. This means that all
directions tangential to the D-brane are compact, and hence the D-brane
has finite mass. {}From the point of view of the full space-time, this
D-brane describes a D-$p$ brane for some $p\ge 1$, with $(p-1)$ directions
wrapped on an internal $(p-1)$ cycle of $\MM$, and one direction
wrapped on the circle of radius $R$ labelled by $x$. On the other hand
from the point of view of an observer who only sees the (2+1) dimensional
space-time labelled by $(x,y,x^0)$, this system corresponds to a D1-brane
wrapped on a circle of radius $R$. {}From now on we shall refer to this
system as the D1-brane or the D-string; with its tension defined as the
total
energy
per unit length along $x$. Of course, an ordinary D-string will be a
special case of
this
system, obtained by putting Dirichlet boundary condition on all the
fields associated with the coordinates on $\MM$. 

The dynamics of an open string with ends on this D-brane is described by a
boundary conformal field theory of central charge 26, which is a direct
sum of the boundary conformal field theories associated with the fields
$X$, $Y$, $X^0$  and the manifold $\MM$. 
We shall denote by
CFT($X$), CFT($Y$) and CFT($X^0$) the boundary conformal field theories
(each
with central charge 1) associated with the fields $X$, $Y$ and $X^0$
respectively, and by CFT$({\cal M})$ the boundary conformal field theory with
central
charge 23 associated with the manifold $\MM$. We also define
\be \label{edefcftp}
\hbox{CFT}' = \hbox{CFT}(Y)\oplus \hbox{CFT}(X^0)\oplus \hbox{CFT}(\MM)\, ,
\ee
so that CFT$'$ has central charge 25. We denote by  $L_n^X$ and $L_n'$ the
Virasoro
generators of CFT($X$) and CFT$'$ respectively. If we denote by
$L_n^{ghost}$ the Virasoro generators of the ghost system, then the
total Virasoro generators of the system will be given by
$L_n=L_n^{ghost}+L_n^X+L_n'$.

The compact direction $x$ corresponds to the direction in which we shall
eventually form the lump. If we follow the normalization convention of
ref.\cite{9912249}, then the tension $\TT_1$ of the D-string described
above is related
to the coupling constant $g_o$ of the open string field theory describing
the wrapped D-string by the relation:
\be \label{eocoup}
2 \pi R \TT_1 = {1 \over 2\pi^2 g_o^2}\, .
\ee
In this normalization convention, a time independent string field
configuration represented by a state $|\Phi\rangle=\Phi(0)|0\rangle$
in the Hilbert space of first quantized string theory, will have a
potential 
\be \label{epot}
\hbox{Potential} = - S(\Phi) = 
{1\over g_o^2} \VV(\Phi) = 2 \pi R \TT_1 \cdot 2 \pi^2
\VV(\Phi)\, ,
\ee
where
\be \label{evph}
\VV(\Phi) = {1\over 2} \langle \,\Phi, Q \Phi \, \rangle 
+ {1\over 3}  \langle \,\Phi , \Phi * \Phi\, \rangle \, .
\ee
Here $Q$ denotes the BRST charge, $\langle , \rangle$ denotes BPZ inner
product between two states, and $*$ denotes the $*$-product of Witten's
open bosonic string field theory~\cite{WITTENBSFT}.

A basis of states in CFT(X) is obtained by acting on
$e^{inX/R}(0)|0\rangle$ with 
the oscillators $\alpha^X_{-m}$ of $X$. It follows
by
a simple counting argument that an alternate basis can be formed out of
the Verma module, containing states obtained by acting on
$e^{inX/R}(0)|0\rangle$ with the operators
$L^X_{-m}$, {\it as long as these states are all linearly independent}.
This is the case if there are no null states in the spectrum. The
condition for the appearance of a null state is given by~\cite{null},
\be \label{enull}
{n^2 \over R^2} = {(p-q)^2 \over 4}\quad \to \quad 
{n \over R} = {(p-q) \over 2}\,,
\ee
where $p$ and $q$ are integers.
Since $n$ is an integer, we can avoid null states for $n\ne 0$ with an
appropriate choice of $R$. Even if we work with a value of $R$ for which
there are null states, the choice of basis described above is good 
below the level where the first null state appears.
{}From now on we shall
restrict our analysis to situations where this choice of basis based on
Verma module is good. In fact our explicit work in later
sections will be based on $R$ values that are not rational, and
thus there will be no null states for $n\not=0$.

\medskip  
For $n=0$, however, there are null states and hence the basis of
states obtained by applying $L^X_{-m}$ on $|0\rangle$ is not complete. 
For example, 
$L^X_{-1}|0\rangle$ is null, and this requires us to explicitly include the 
primary state $\alpha^X_{-1}|0\rangle$ in the basis. 
There are further null
states in the Verma module over $\alpha^X_{-1}|0\rangle$, and hence there
are
new primary states at higher level which must be explicitly included in
the basis. Let us denote by $\{|\vp^i_e\rangle
=\vp^i_e(0)|0\rangle_{X}\}$
and $\{|\vp^i_o\rangle=\vp^i_o(0)|0\rangle_X\}$ the set of zero momentum
primary states which
are respectively even and odd under the reflection $X\to -X$. The complete
basis of zero momentum states in CFT(X) is obtained by acting on 
$\{|\vp^i_e\rangle\}$
and $\{|\vp^i_o\rangle\}$ with $L^X_{-n}$'s, and removing the
null states.

A generic string field configuration is represented by an arbitrary state
in the Hilbert space $\HH$ of ghost number one in the  combined matter,
ghost conformal field theory. We now claim that in order to discuss a
lump along the
$x$ coordinate, we can restrict the string field $|\Phi\rangle$
to a subspace
$\widehat{\cal
H}$ of $\HH$, built by acting with the oscillators
\begin{equation}
\label{list}
\{ L_{-1}^X, L_{-2}^X, \cdots  ;  L'_{-2}, L'_{-3}, \cdots ;
c_{1}, c_{-1}, c_{-2} , \cdots ; b_{-2},b_{-3}, \cdots \}  
\end{equation}
on the following primary states:  
\begin{itemize}

\item The zero momentum even primaries $\vp^i_e(0)|0\rangle$
(and removing the null states), and,

\item The Fock
vacuum states of the form
\begin{equation}
\label{vacua}
\cos{\left( {n \over R} X(0) \right)}\ket{0} = {1\over 2}
\Bigl( e ^{inX(0)/R} +  e ^{-inX(0)/R}\Bigr) \ket{0} 
= {1\over 2}  \Bigl( \,\ket{{n\over R}} + \ket{-{n\over R}} \Bigr) \quad 
n\ne 0\,,
\end{equation}

\end{itemize}
where $\ket{0}$ is the SL(2,R) vacuum of the combined matter, ghost
conformal field theory. A few points should be made. The Virasoro
operator $L'_{-1}$ is not required for it kills the above primary
states (this is not the case for $L^X_{-1}$). $b_{-1}$ and $b_0$ also
annihilate the vacuum $|0\rangle$, and hence have been omitted from the
list. We have not
included the oscillator $c_0$ because we work in the Siegel gauge, where
all states must be annihilated by $b_0$. Finally we can restrict
ourselves to states of even twist \cite{9705038}. This simply requires that 
the eigenvalue of the number operator $\wh N$ must be odd
(same as that for the tachyon).

In order to show that the above is a consistent truncation
of the string field, one must show that there is no 
term in the action that couples a single state in  
$({\cal H} - \widehat {\cal H})$ to a state in $\widehat{\cal H}$ via the 
quadratic term, or to a pair of states in $\wh \HH$ via the interaction term.
This is readily done by listing the states in $({\cal H} - 
\widehat {\cal H})$.
We carry along all ghost oscillators and classify the states
by
their behavior under the matter  operators. In this
way we get the 
following disjoint sets:
\begin{itemize}

\item States with nonzero momentum $k_0$ along $X^0$.

\item States obtained by acting with the oscillators in \refb{list} 
on Fock vacua of the type $\sin ({nX(0)\over R})\ket{0}$, or on a state of
the
form
$\vp^i_o(0)|0\rangle$.

\item  States obtained by acting with \refb{list}  
on states that  (i) have $k_0=0$,  (ii) are non-trivial primaries  of
CFT$'$ (of
dimension greater than zero, by unitarity), and (iii) 
are CFT(X) primaries. 

\end{itemize}
It is manifest by momentum 
conservation that a state in the first set 
cannot couple to states in $\widehat{\cal H}$.
The symmetry $X \to -X$ of CFT(X) insures that a state
in the second set also cannot couple
to states in $\widehat{\cal H}$. The same is true for
the last set as  Virasoro 
Ward identities can be used to show that a correlator
involving two states in  $\widehat{\cal H}$ and a state
in the last set is proportional to the one point function
of the CFT$'$ primary in question. Since this primary must
have dimension greater than zero, its one point function
vanishes. This completes our justification for the
use of $\widehat{\cal H}$.

Since the choice of basis described above requires the use of the basis
$\{|\vp^i_e\rangle\}$, it will be useful to determine at which level the
first zero momentum primary (other than the vacuum state) appears. For
this we can
compare the full partition function of CFT(X) for states even under $X\to
-X$
\be \label{eparteven}
Z_{even}(q) \equiv Tr_{even}(q^{L_0^X - {1\over 24}}) ={1\over 2}\,
q^{-{1\over 24}}\Big(
\prod_{n=1}^\infty {1 \over 1 - q^n} +  
\prod_{n=1}^\infty {1 \over 1 + q^n} \Big)
\ee
with the Virasoro character for $(c=1,h=0)$~\cite{null}, 
\be \label{echarac}
\chi_{c=1,h=0}(q) = q^{-{1\over 24}} \prod_{n\ge 2}{1\over 1 - q^n}\, .
\ee
It can be easily checked that
\be \label{ediff}
Z_{even}(q) - \chi_{c=1,h=0}(q) = q^{-{1\over 24}} (q^4 +O(q^5))\, .
\ee
Thus the first non-trivial primary $|\vp^1_e\rangle$ even under $X\to -X$
appears at level four.  Indeed, the available descendents at this level,
$L_{-4}^X \ket{0}, L_{-2}^XL_{-2}^X\ket{0}$, do not suffice to represent
the nonvanishing 
$X$-even  states $\alpha_{-3}^X\alpha_{-1}^X\ket{0}, 
\alpha_{-2}^X\alpha_{-2}^X\ket{0},
(\alpha_{-1}^X)^4\ket{0}$. 

\subsection{Mass of the lump}

To begin with, the wrapped D-$p$ brane, which we have been calling a D-1
brane wrapped on a circle of radius $R$, has mass $2\pi R \TT_1$, where
$\TT_1$, as defined earlier, is the tension of this D1-brane.
We want to compute the mass of the system in a
situation where the tachyon 
field on this D$1$-brane develops a lump along a circle of 
radius $R$ (this direction is represented
by the world sheet field $X$). 
If we denote by $\vec T$ the multicomponent string field
configuration
on the D1 brane, restricted to $\wh \HH$,  then, using eq.\refb{epot}, the
rest mass energy plus
potential energy of the D1 brane
stretched on the circle can be written as
\begin{equation} \label{energy1}
E(D1) =  {\cal T}_1 (2\pi R )\,\bigl( 1 + 2 \pi^2 \VV(\vec T)
\bigr)
\end{equation}
where $\VV$ has been defined in eq.\refb{evph}.
Before condensation,
$\vec T =0$ and
$\VV(\vec T) =0$, and thus the energy formula correctly reproduces the
mass
of
the D1-brane. Recall that for the nontrivial 
translationally invariant vacuum
$\vec T_{vac}$, one expects $\VV(\vec T_{vac}) = -1/(2\pi^2)$ and  the
energy formula correctly gives zero (as the D1 brane has disappeared).
Using $\VV(\vec T_{vac}) = -1/(2\pi^2)$ we can write the energy formula as
\begin{equation} \label{energy2}
E (\hbox{D1}) = \TT_1\, 2\pi R \cdot 2\pi^2\,\bigl(\VV(\vec T) -
\VV(\vec T_{vac}) \bigr)
\end{equation}

The mass of the tachyonic lump solution, represented by the configuration
$\vec T_{lump}$, is obtained by replacing $\vec T$ by $\vec T_{lump}$ on
the right hand sides of eqs.\refb{energy1} or \refb{energy2}.
This tachyonic lump on the D-string (wrapped D-$p$-brane) is conjectured
to
be equivalent to a D0-brane (a wrapped D-$(p-1)$-brane) of mass $\TT_0$.
With
$\alpha'=1$, the ratio of the tension of a D-$p$ brane and a
D-$(p-1)$-brane is $1/(2\pi)$; using this we get,
\be \label{emd0}
\TT_0=2\pi \TT_1\, .
\ee
This gives
\be \label{predict}
r \equiv {E_{lump}\over \TT_0} = 
2 \pi^2 R \,\bigl(\,\VV(\vec T_{lump}) - \VV(\vec
T_{vac})\bigr)\, .
\ee
The predicted answer for this ratio is 1.

This prediction can be tested for various
values of $R$, and independently of the chosen value we must obtain
unity, since the mass of a D0-brane on a circle of radius $R$ does not
depend on $R$. At fixed $R$ and at any level of approximation in the level
expansion 
it is possible to use \refb{predict} in two ways.
We can use that $2\pi^2 \VV(\vec T_{vac})$ at the exact 
vacuum is indeed $-1$ and thus we check how accurately
\begin{equation}
\label{foption}
r^{(1)} \equiv   R \,\bigl( 2 \pi^2 \VV_{(M,N)} ({\vec T}_{lump}) +1
\bigr)
\end{equation}
approaches unity. Here $\VV_{(M,N)}$ is the potential calculated at
the specified level of approximation.  Alternatively we can
use the translationally invariant vacuum that is obtained with
the same level of approximation used to compute the lump.\footnote{The
value of $\VV_{(M,N)}(\vec T_{vac})$ can be read from
refs.~\cite{9912249,0002237}.} This gives
\begin{equation}
\label{soption}
r^{(2)} \equiv   R \,\bigl( 2 \pi^2 \VV_{(M,N)} ({\vec T}_{lump}) -
2\pi^2 \VV_{(M,N)}(\vec T_{vac}) \bigr)\,
\end{equation}
We will find that $r^{(1)}$ approaches unity monotonically
from above as we increase the level of approximation. On the other hand
$r^{(2)}$ provides a more accurate
answer.  

\subsection{Setup and Sample Computations}

Let us now describe explicitly the string 
field we will be using to analyze the bosonic
string lump.  The zero momentum tachyon $\ket{T_0} = c_1 \ket{0}$ 
now becomes the lowest in a family of states
\begin{equation}
\label{tachtow}
\ket{T_n} = \displaystyle{c_1 \cos{\left({n \over R} X(0) \right)}}
\ket{0}\,, \quad  l(T_n) = {n^2\over R^2}
\end{equation}
where $l(T_n)$ denotes the level of $T_n$.
For any given computation
only a finite number of tachyon modes are required. In the zero-momentum
computation the next modes that contribute are $\ket{U_0} = c_{-1}
\ket{0}$
and $\ket{V_0} = L_{-2}^{matt} \ket{0}.$ In view of our remarks
around \refb{vacua} these states actually give rise to three towers 
\begin{eqnarray} \label{uvwtow} 
\ket{U_n} &=& 
\displaystyle{
\,\,c_{-1} \,\,\,\cos{\left({n \over R} X(0)\right)} }\ket{0}
\,, \quad  l(U_n) = 2 + {n^2\over R^2} \,,\nonumber \crbig
\ket{V_n} &=& \displaystyle{c_1 L^X_{-2} \cos{\left({n \over R}
X(0)\right)}} \ket{0}
\,, \quad  l(
V_n) = 2 + {n^2\over R^2}\,,\nonumber \crbig
\ket{W_n} &=& \displaystyle{c_1 L^{\prime}_{-2} \cos{\left({n \over R}
X(0)
\right)}}\ket{0}\,, \quad  l(W_n) = 2 + {n^2\over R^2}\,.
\end{eqnarray}
%\]
In addition to these three towers there is one more, where
the $n=0$ state happens to vanish:
\be \label{ztower}
\ket{Z_n} = \displaystyle{
c_1 \, L_{-1}^X L_{-1}^X \cos{ \left({n \over R}
X(0) \right)}}\ket{0}\,, \quad n\geq 1, \quad   l(Z_n) = 2 + {n^2\over
R^2}\,.
\ee
No new fields or towers arise until level four, and for the
purposes of the present paper we shall not carry computations that 
far.  Therefore we will use the string field
\begin{eqnarray}
\ket {\vec T} &=& \,\,\,\, t_0 \ket{T_0} + t_1 \ket{T_1}+ t_2 \ket{T_2}
+\cdots \nonumber\crbig
&& + \,u_0 \ket{U_0} + u_1 \ket{U_1} +\cdots\crbig\nonumber
&& + \,v_0 \ket{V_0} + v_1 \ket{V_1} +\cdots\crbig\nonumber
&& + \,w_0 \ket{W_0} + w_1 \ket{W_1} +\cdots\crbig\nonumber
&& + \,z_1 \ket{Z_1}   +\cdots\nonumber
\end{eqnarray}
Which fields and which interactions must
be kept for any fixed level computation depends on the
chosen radius, and this will be discussed in the following
sections.  We conclude here with some basic comments 
about the evaluation of the potential (or the action) for
a string field of the above type.

This is simply the evaluation of $\VV(\vec T)$ as given in \refb{evph}
\begin{equation}
\VV(\vec T) = {1\over 2} \langle \,\vec T, Q \vec T \,\rangle 
+ {1\over 3}  \langle \,\vec T , \vec T * \vec T\, \rangle \,.
\end{equation}
We work in units where 
$\alpha^{\prime} = 1$.
The stress tensor for the compact coordinate $X$ is
$T_X = - {1 \over 4}  \partial X \,\partial X$
with 
$ X(z) X(w) \sim -2 \ln (z-w)$, 
$T(z) e^{i p\cdot X(w)}\sim 
{p^2 \over (z-w)^2} e^{i p \cdot X(w)}$
and $ 
e^{i p_1 \cdot X(z)}  e^{i p_2 \cdot X(w)} = 
(z-w)^{2 p_1 \cdot p_2} e^{i p_1 \cdot X(z)+i p_2 \cdot X(w)} 
$, where $z$ and $w$ are coordinates on the real line with $z>w$.
With these conventions
$L_0 \ket{p} = L_0 e^{i p \cdot X(0)} \ket{0} = p^2 \ket{p}$.
In addition, the inner product is normalized as
\begin{equation}
\bra{{n\over R}} \, c_{-1} c_0 c_1 \ket{{m\over R}} = \delta_{n,m}
\end{equation}
Consider, for example contributions from the tachyon tower
to the action. By momentum conservation all kinetic terms
must be diagonal. Using \refb{vacua} we see that the 
contribution from $t_n$ ($n\geq 1$) to $V$ is
\begin{equation}
{1\over 2}\,{t_n\over 2}\, {t_n\over 2}  \Bigl( \,\bra{ -{n\over R}} +
\bra{{n\over R}} \Bigr) c_{-1} c_0 L_0  c_1 \Bigl( \,\ket{{n\over R}}
+\ket{-{n\over R}}\Bigr) 
\end{equation}
By momentum conservation there are two cross terms that do not
vanish and give identical contributions.  We thus get 
\begin{equation}
{1\over 4}\,t_n^2  
\bra{{n\over R}} c_{-1} c_0 \Bigl( -1  + {n^2\over R^2}\Bigr) c_1 
\,\ket{{n\over R}}  = -{1\over 4}\Bigl( 1  - {n^2\over R^2}\Bigr) \,t_n^2  
\end{equation}
For $t_0$ the normalization factor differs by a factor of two.  All this
together gives us that the quadratic terms are
\begin{eqnarray}
\label{quadtach}
\VV(t_0, t_1, t_2, \cdots)^{(2)} &=& -{1\over 2} t_0^2  -{1\over 4} 
\sum_{n=1}^\infty 
 \Bigl( 1  - {n^2\over R^2}\Bigr) \,t_n^2\nonumber \crbig
&=& -{1\over 2} t_0^2 
 -{1\over 4}\Bigl( 1  - {1\over R^2}\Bigr) \,t_1^2
-{1\over 4}\Bigl( 1  - {4\over R^2}\Bigr) \,t_2^2 + \cdots
\end{eqnarray}
We will use the first tachyon harmonic $t_1$ to drive the unstable
vacuum into the lump solution. Note that $t_1$ is tachyonic whenever
$R>1$.  We will choose different values of $R>1$ to examine how the lump
forms.
As $R$ increases, more and more tachyon harmonics become tachyonic.

It is not difficult to compute the interactions of the various
tachyon harmonics. One can use the oscillator expressions for the 
states and contract them against 
the 3-string vertex bra $\bra{V_{123}}$~\cite{gross}.
Alternatively
one can use the conformal field theory definition~\cite{LPP}
\be \label{e2} \langle \vec T , \vec T * \vec T \rangle
\equiv \langle h_1 \circ T(0) h_2 \circ 
T(0) h_3 \circ T(0) \rangle \,\, .
\ee  
where $T(0)$ denotes the vertex operator associated
to the state $\ket{\vec T}$. Here $h_1$, $h_2$ and $h_3$
are a set of familiar conformal transformations 
reviewed in~\cite{9911116}. For illustration purposes consider
three tachyon harmonics $t_n, t_m$ and $t_{n+m}$, with
$n\not= m \not=0$. Such fields contribute to $\VV$ the following
interaction
\be
{1\over 3} \cdot 6 \cdot {t_n\over 2} 
\cdot {t_m\over 2} \cdot{t_{n+m}\over 2} 
\cdot 2\, \langle \, h_1\circ  (c e^{inX\over R}) (0)\,\, h_2 \circ 
(c e^{imX\over R}) (0)\,\, h_3 \circ (c e^{-i(n+m)X\over R})(0) \rangle
\ee
The factor $(1/3)$ is in the definition of $\VV$. The factor of $6$
appears
because this is the number of ways three different fields can be assigned
to the three punctures in the disk. Then come the fields, and then a 
factor of two, as there are two momentum conserving combinations giving
equal contributions. Evaluation of the above gives
\be \label{ekdef}
{1\over 2} \, t_n\, t_m\, t_{n+m}\, 
K^{3 -{1\over R^2}(n^2 + m^2 + (n+m)^2)}\,,
\quad K \equiv {3\sqrt{3}\over 4}
\ee
Slightly different combinatorics are required for
terms of the form $t_0t_n^2$ and $t_n^2 t_{2n}$.
Combining all such terms together we obtain\
\begin{eqnarray}
\label{cubictach}
\VV(t_0, t_1, \cdots)^{(3)} &=& {1\over 3} K^3 t_0^3 + {1\over
2}\,\sum_{n=1}^\infty t_0 \,t_n^2\,\, K^{3- {2n^2\over R^2}}
+ {1\over 4}\,\sum_{n=1}^\infty
t_n^2 \,t_{2n}\,\, K^{3- {6n^2\over R^2}}\nonumber \crbig
&& \,\, + {1\over 2} \sum_{n \geq 1}^\infty \sum_{m> n}^\infty\,
t_n\, t_m\, t_{n+m}\, K^{3 -{2\over R^2}(n^2 + m^2 + nm)}\,\,.
\end{eqnarray} 
Equations \refb{quadtach} and \refb{cubictach} give the complete
potential for the tachyon tower.

\sectiono{Calculating the action in the Level Expansion for $R=\sqrt{3}$}
\label{s2}

In this section we will consider different truncation levels to calculate 
the lump tension. For this we will write explicitly the action at 
different levels. Though we will work with a fixed radius $R = \sqrt{3}$, 
all our equations will contain $R$ as a variable for further use.
Once we know the action we can solve the equations of motion numerically
for
the one-lump 
solution 
by giving a nonzero initial value to $t_1$. At the end of the section, we 
will be able to study the convergence of our level truncation scheme by 
using both (\ref{foption}) and (\ref{soption}).

\begin{table}
\begin{center}\def\st{\vrule height 3ex width 0ex}
\begin{tabular}{|l|l|} \hline \label{fields}
Level & Fields \st\\[1ex] \hline \hline
0 & $t_0$ \st\\[1ex] \hline
1/3 & $t_1$ \st\\[1ex] \hline
4/3 & $t_2$ \st\\[1ex] \hline
2 &  $u_0, \ v_0, \ w_0$ \st\\[1ex] \hline
7/3 & $u_1, \ v_1, \ w_1, \ z_1$ \st\\[1ex] \hline
3 & $t_3$ \st\\[1ex] \hline
\end{tabular} 
\end{center}
\caption{The list of fields appearing at various 
levels when $R = \sqrt{3}$.}
\label{entable} 
\end{table}

We will do these calculations at levels $(1/3, 2/3)$, $(4/3, 8/3)$, $(2,
4)$, 
$(7/3, 14, 3)$ and $(3, 6)$. This will require the fields listed in
Table~\ref{entable}
with their respective levels (using (\ref{tachtow}), (\ref{uvwtow}) and 
(\ref{ztower})). 
In order to study the truncation method at various levels, we define 
$V(m, n)$ to be the part of the whole potential 
satisfying the three following conditions:

\begin{enumerate}
\item All terms in $V(m, n)$ have level $n$.
\item All terms in $V(m, n)$ contain only fields of level smaller than 
or equal to $m$.
\item All terms in $V(m, n)$ contain at least one field of level $m$.
\end{enumerate}

This definition ensures that various $V(m, n)$'s are disjoint 
(i.e. $V(m, n)$ and $V(m^{\prime}, n^{\prime})$ don't contain common terms 
for $(m, n) \neq (m^{\prime}, n^{\prime})$).
It now follows that the total potential at level $(M, N)$ is given by
\beq
{\cal V}_{(M, N)} = \sum_{m \leq M} \sum_{n \leq N} V(m, n)
\label{rectangle}
\eeq

We shall now compute $V(m, n)$ for $m \leq 3$
and 
$n \leq 6$.
Though here we will restrict ourselves to levels $(M,N)$ of the form $(M,
2M)$, 
eq.(\ref{rectangle}) and the results for $V(m,n)$ given below can be used
to
construct the potential $\VV_{(M,N)}$ for 
arbitrary level $(M, N)$ as long as 
$M \leq 3$ and $N \leq 6$. We shall first
list all possible terms appearing in 
each $V(m,n)$ consistent with momentum
conservation, separating
the quadratic and cubic terms. 
We then use the methods described in section \ref{s1} to 
explicitly calculate the coefficients of each possible term in the $V(m,
n)$'s. 

The  list of interactions that must
be computed is generated conveniently with the help of
the following
function:
\begin{eqnarray}
\label{genfunc}
{\cal Z} (x,y,s) &\equiv& \prod_{n=0}^\infty 
\Bigl\{ \Bigl (1 - t_n x (y^n + y^{-n}) s^{n^2/R^2} \Bigr)
\Bigl (1 - u_n x (y^n + y^{-n}) s^{2+ n^2/R^2} \Bigr)\nonumber \\
&&\qquad \Bigl (1 - v_n x (y^n + y^{-n}) s^{2+ n^2/R^2} \Bigr)
 \Bigl (1 - w_n x (y^n + y^{-n}) s^{2+ n^2/R^2} \Bigr)
\nonumber \\
&&\qquad \Bigl (1 - z_{n+1} x (y^{n+1} 
+ y^{-n-1}) s^{2+ (n+1)^2/R^2} \Bigr) \cdots
\Bigr\}^{-1}
\end{eqnarray} 
Here the formal variables $x, y$ and $s$ are used to count
number of fields, momentum, and level, respectively. If we write 
\be
{\cal Z} (x,y,s) = \sum_{m,n}  {\cal Z} (m,n,s) \,x^m y^n\,. 
\ee
The momentum conserving cubic interactions appear in ${\cal Z} (3,0,s)$
and an expansion in $s$ gives
\be
{\cal Z} (3,0,s) = \sum_l {\cal Z} (l) \, s^l\,. 
\ee 
Let $\{ \psi^i \}$ denote the complete set of 
modes $(t_n, u_n , \cdots )$. Then ${\cal Z}(l)$ 
has an expression of the form ${\cal Z}(l) 
\sim \sum a_{ijk} \psi^i \psi^j\psi^k$
where each $a_{ijk}$ is an integer. If $a_{ijk} \not=0$ the
interaction $\psi^i\psi^j\psi^k$ must be included in the
level $l$ contribution to the potential. Thus
${\cal Z}(l)$ supplies the complete list of momentum
conserving cubic interactions of level $l$. 
When useful, we split by hand the terms in ${\cal Z}(l)$ to obtain the
possible terms which appear in various $V(m, l)$'s.

The list of all terms for the various $V(m,n)$'s
with $n\leq 6$ (and $R=\sqrt{3}$) are given in Table~\ref{mainlistt}. 

\begin{table}
\begin{center} \def\st{\vrule height 3ex width 0ex}
\begin{tabular}{|l||l|l|} \hline

& Quadratic terms & Cubic terms \st\\[1ex] \hline \hline

$V(0, 0)$ & $t_0^2$ & $t_0^3$ \st\\[1ex]\hline \hline

$V(1/3, 2/3)$ & $t_1^2$ & $t_0 t_1^2$ \st\\[1ex] \hline \hline

$V(4/3, 2)$ & & $t_1^2 t_2$ \st\\[1ex] \hline
$V(2, 2)$ & &  $t_0^2 u_0, \ t_0^2 v_0, \ t_0^2 w_0$ \st\\[1ex] 
\hline \hline

$V(4/3, 8/3)$ & $t_2^2$ &
 $t_0 t_2^2$ \st\\[1ex] \hline
$V(2, 8/3)$ & & $t_1^2 u_0, \ t_1^2 v_0, \ t_1^2 w_0$ \st\\[1ex] \hline
$V(7/3, 8/3)$ & & $t_0 t_1 u_1, \ t_0 t_1 v_1, \ t_0 t_1 w_1, \ t_0 t_1
z_1$ 
\st\\[1ex] \hline \hline

$V(2, 4)$ & $u_0^2, \ v_0^2, \ w_0^2$ & $t_0 u_0^2, \ t_0 v_0^2, 
\ t_0 w_0^2, \ 
t_0 u_0 v_0, \ t_0 u_0 w_0, \ t_0 v_0 w_0$ \st\\[1ex] \hline
$V(7/3, 4)$ & & $t_1 t_2 u_1, \ t_1 t_2 v_1, \ 
t_1 t_2 w_1, \ t_1 t_2 z_1$ \st\\[1ex] \hline \hline

$V(2, 14/3)$ & &$t_2^2 u_0, \ t_2^2 v_0, \ t_2^2 w_0$ \st\\[1ex] \hline
$V(7/3, 14/3)$ &  $u_1^2, \ v_1^2, \ w_1^2, \ z_1^2, $ & $t_0 u_1^2, \ t_0
v_1^2, 
\ t_0 w_1^2, \ t_0 u_1 v_1, \ 
t_0 u_1 w_1, \ t_0 v_1 w_1$,  \st\\ 
& $v_1 z_1$ & $t_1 u_0 u_1, \ t_1 u_0 v_1, \ 
t_1 u_0 w_1, \ t_1 v_0 u_1, \ t_1 v_0 v_1$,  \st\\ 
& & $t_1 v_0 w_1, \ 
t_1 w_0 u_1, \ t_1 w_0 v_1, \ t_1 w_0 w_1$, \st\\ 
& & $t_0 z_1^2, \ t_0 u_1 z_1, \ t_0 v_1 z_1, \ 
t_0 w_1 z_1$, \st\\ 
& & $t_1 u_0 z_1, \ t_1 v_0 z_1, \ t_1 w_0 z_1$
\st\\[1ex]\hline
$V(3, 14/3)$ & & $t_1 t_2 t_3$  \st\\[1ex]\hline \hline

$V(2, 6)$ & & $u_0^3, \ v_0^3, \ w_0^3, \ u_0^2 v_0, \ u_0^2 w_0, \ u_0
v_0^2, 
\ u_0 w_0^2, \ v_0^2 w_0$ \st\\ 
& & $v_0 w_0^2, \ u_0 v_0 w_0$, \st\\[1ex] \hline
$V(7/3, 6)$ & & $t_2 u_1^2, \ t_2 v_1^2, \ t_2 w_1^2, 
\ t_2 u_1 v_1, \ t_2 u_1 w_1, \ t_2 v_1 w_1, $ \st\\ 
& & $t_2 z_1^2, \ t_2 u_1 z_1, \ 
t_2 v_1 z_1, \ t_2 w_1 z_1$ \st\\[1ex] \hline
$V(3, 6)$ & $t_3^2$ & $t_0 t_3^2$ \st\\[1ex] \hline 
  
\end{tabular}
\end{center}
\caption{Quadratic terms and
interactions appearing at various 
levels when $R = \sqrt{3}$.}
\label{mainlistt} 
\end{table}

\subsection{The terms in the potential}
\noindent 
The explicit interactions corresponding
to the various terms appearing in the table
will be listed here. With
$K = {3 \sqrt 3\over 4}$, 
as in eq.\refb{ekdef},
we have at the lowest level:  
%\[
\begin{eqnarray} %{lcl} 
\label{v0}
V(0, 0) & = & 
\displaystyle{- \half\, t_0^2 + {1 \over 3} K^3 t_0^3} \,.%\\
\end{eqnarray}
%\]
\noindent At first nontrivial level we have:
%\[
\begin{eqnarray} \label{v23} %{lcl}
V(1/3, 2/3) & = &
\displaystyle{- {1 \over 4} \left(1 - {1 \over R^2} \right) t_1^2
 + {1 \over 2} K^{3 - 2/R^2} t_0 t_1^2}\,. 
\end{eqnarray}
%\]
\noindent At level 2 we have:
%\[
\begin{eqnarray} \label{v2} % {lcl}
V(4/3, 2) & = &
\displaystyle{
{1 \over 4} K^{3 - 6/R^2} t_1^2 t_2} \nonumber \\[2ex]
V(2, 2) & = &
\displaystyle{ {K\over 32}\,t_0^2\, \Bigl(  22\, u_0
 -  5 \, (v_0 + 25 w_0) \Bigr)\,.}
\end{eqnarray}
\noindent At level 8/3 :
\begin{eqnarray} \label{v83} % {lcl}
V(4/3, 8/3) & = & \displaystyle{ 
- {1 \over 4} \left(1 - {4 \over R^2} \right) t_2^2
 + 
{1 \over 2} K^{3 - 8/R^2} t_0 t_2^2} \nonumber \\[3ex]
V(2, 8/3) & = &
\displaystyle{ \, K^{1-2/R^2} t_1^2\, \Bigl(  \,
{11  \over 32}  u_0 + 
\half \left( {1 \over R^2} - {5 \over 32} \right) \, v_0 - 
{125 \over 64} \, w_0\Bigr)}  \nonumber \\[3ex]
V(7/3, 8/3) & = &
\displaystyle{ {1 \over 32} K^{1 - 2/R^2} t_0 t_1 
\left( 22 u_1 - \left( 5 + {16 \over R^2} \right) v_1 - 125 w_1 
+  \left( {-44\over R^2} + 
{ 32 \over R^4} \right) z_1 \right)}\,.\nonumber \\ 
\end{eqnarray}
\noindent At level 4:
\begin{eqnarray} \label{v4} % {lcl}
V(2, 4) & = & 
\displaystyle{- \half\, u_0^2
 + {1 \over 4} (v_0^2 + 25 w_0^2) 
+ K \left\{ {1 \over 576}\, t_0 \left( 76 \, u_0^2 + 
179 \, v_0^2 + 9475 \, w_0^2 \right) \right.} \nonumber 
\\[3ex] %\crbig
& &
\displaystyle{ \left.
+ {625 \over 864} \,\,t_0 v_0 w_0 -
{55 \over 432}\,\, t_0 u_0 (v_0 + 25 w_0) \right\} } \nonumber \\[3ex]
V(7/3, 4) \hskip-6pt & = &
\hskip-6pt\displaystyle{
 {1 \over 64} K^{1 - 6/R^2} t_1 t_2 \left( 22 u_1 - \left( 5 - 
{48 \over R^2} \right) v_1 \hskip-3pt- 125 w_1 + 
\left(\hskip-4pt - {44\over R^2} + {288 \over R^4} \right) 
\hskip-2pt z_1 \right)}\,.\nonumber\\ 
\end{eqnarray}
\noindent At level 14/3:
\begin{eqnarray} \label{v143} 
V(2, 14/3) & = &
\displaystyle{
 {1 \over 64} K^{1 - 8/R^2} t_2^2 \left( 22 u_0
- \left( 5 - {128 \over R^2} \right) v_0 - 125 w_0 \right) } \nonumber 
\\[3ex]
V({7\over 3}, {14\over 3}) & = & 
\displaystyle{ 
{1 \over 8} \left( 1 + {1 \over R^2} \right) \left( -2 u_1^2 +
\left( 1 + {8 \over R^2} \right) 
v_1^2 + 25 w_1^2 
+ \left( {8 \over R^2} + {16 \over R^4} \right) z_1^2
+ {24 \over R^2} v_1 z_1 \right) } \nonumber \crbig
& & 
\displaystyle{+ K^{1 - 2/R^2} \left\{ {19 \over 288} t_0 u_1^2 
+ {1 \over 3456}  \left( 
537 + {8864 \over R^2} + {256 \over R^4} \right)  t_0 v_1^2 + {28425 
\over 3456}  t_0 w_1^2 \right.}
\nonumber \crbig
& & 
\displaystyle{
 - {11 \over 864} t_0 u_1 
\left( \left( 5 + {16 \over R^2} 
\right) v_1 + 125 w_1 \right) + {125 \over 1728}  \left( 5 + {16 \over
R^2} \right) 
t_0 v_1 w_1 } \nonumber \crbig
& & 
\displaystyle{
+ {19 \over 144}  t_1 u_0 u_1 - {11 \over 864}  t_1 u_0 \left( \left( 5 +
{16 \over R^2}
\right) v_1 + 125 w_1 \right)} \nonumber \crbig
& &
\displaystyle{
- {11 \over 864} \left( 5 - {32 \over R^2} \right) 
t_1 v_0 u_1 + {1 \over 1728} \left( 
537 + {944 \over R^2} - {512 \over R^4} 
\right) t_1 v_0 v_1} \nonumber \crbig
& &
\displaystyle{ 
+ {25 \over 1728}  \left( 25 - {160 \over R^2} 
\right)  t_1 v_0 w_1 - {1375 \over 864} 
t_1 w_0 u_1 + {25 \over 1728} \left(
25 + {80 \over R^2} \right)  t_1 w_0 v_1} \nonumber \crbig 
& & \displaystyle{
+ {28425 \over 1728}  t_1 w_0 w_1 + {1 \over 216}  { 1 \over R^2} 
\left( 384 + {1145 \over R^2} + 
{336 \over R^4} + {64 \over R^6} \right) t_0 z_1^2} \nonumber \crbig
& & 
\displaystyle{
+ {11 \over 864} {1 \over R^2}\left( -44 + {32 \over R^2} \right)  
t_0 u_1 z_1 + {1 \over 432} \left( {2359 \over R^2} + 
{1672 \over R^4} - {128 \over R^6} \right) t_0 v_1 z_1} \nonumber \crbig
& & 
\displaystyle{ 
+ {125 \over 432} {1 \over R^2} \left( 11 - 
{8 \over R^2} \right) \left( t_0 w_1 z_1 + t_1 w_0 z_1 \right)
} \nonumber \crbig
& & 
\displaystyle{ \left.
+ {1 \over 864} {1 \over R^2} \left( 11 \left( -44 + {32 \over R^2}
\right) 
t_1 u_0 z_1 + \left( 2158 - {2832 \over R^2} + {512 \over R^4} \right)  
t_1 v_0 z_1 \right) \right\}} \nonumber \\[3ex]
V(3, 14/3) & = & \displaystyle{ 
\half K^{3 - 14/R^2} t_1 t_2 t_3}\,.
\end{eqnarray}
\noindent And at level 6:
\begin{eqnarray} \label{v6} 
V(2, 6) & = & \displaystyle{
K \left\{ {1 \over 144} u_0^3 + {8321 \over 93312} 
v_0^3 - {219775 \over 10368} w_0^3  
- {95 \over 7776} u_0^2 
\left( v_0 + 25 w_0 \right) \right. } \nonumber \crbig
& & \displaystyle{
+ {1969 \over 15552} u_0 v_0^2 + 
{104225 \over 15552} u_0 w_0^2 - 
{22375 \over 31104} v_0^2 w_0} 
\displaystyle{\left. 
- {47375 \over 31104} v_0 w_0^2 
+ {6875 \over 23328} u_0 v_0 w_0 \right\} } \nonumber \\[3ex]
V(7/3, 6) & = & \displaystyle{
K^{1 - 6/R^2} \left\{ {19 \over 576} t_2 u_1^2 + 
{1 \over 2304} \left( 179 - {1696\over R^2} 
+ {768 \over R^4} \right) t_2 v_1^2 
+ {9475 \over 2304} t_2 w_1^2 \right. } \nonumber \crbig 
& & \displaystyle{
- {11 \over 1728} \left( 5 - {48 \over R^2} \right) t_2 u_1 v_1 
- {1375 \over 1728} t_2 u_1 w_1
+ {1 \over 72} \left( {625 \over 48} - {125 \over R^2}  
 \right) t_2 v_1 w_1 } \nonumber \crbig
& & \displaystyle{
+ { 1 \over 144} \left( - {128 \over R^2} + {723 \over R^4} 
- {2064 \over R^6}  + {1728 \over R^8} \right) t_2 z_1^2 + 
{11 \over 432}
\left( {-11 \over R^2} + {72 \over R^4} \right) 
t_2 u_1 z_1} \nonumber \crbig
& & \displaystyle{ \left.
+ { 1 \over 288} \left( - {67 \over R^2}  - {808 \over R^4} 
+ {1152 \over R^6} \right) t_2 v_1 z_1 
+ { 125 \over 864} \left( {11 \over R^2} - {72 \over R^4} 
\right) t_2 w_1 z_1 \right\} } \nonumber \\[3ex]
V(3, 6) & = & \displaystyle{
{1 \over 4 } \left( -1 + {9 \over R^2} \right) t_3^2
+ \half K^{3 - 18/R^2} t_0 t_3^2} \,.
\end{eqnarray}

\subsection{Potentials at various truncation levels and mass calculations}

{}From these formulae one can construct 
the potentials at various truncation 
levels using (\ref{rectangle}). As we will use them, we give below 
the explicit sums for 
${\cal V}_{(1/3, 2/3)}$, ${\cal V}_{(4/3, 8/3)}$, ${\cal V}_{(2, 4)}$, 
${\cal V}_{(7/3, 14/3)}$ and ${\cal V}_{(3, 6)}$:
\beq \label{vtot}
\begin{array}{rcl}
{\cal V}_{(1/3, 2/3)} & = & V(0, 0) + V(1/3, 2/3) \crbig

{\cal V}_{(4/3, 8/3)} & = & {\cal V}_{(1/3, 2/3)} +  V(4/3, 2) + V(4/3,
8/3)
\crbig

{\cal V}_{(2, 4)} & = & {\cal V}_{(4/3, 8/3)} + V(2, 2) + V(2, 8/3) + V(2,
4) 
\crbig

{\cal V}_{(7/3, 14/3)} & = & {\cal V}_{(2, 4)} + V(7/3, 8/3) + V(7/3, 4) + 
V(2, 14/3) + V(7/3, 14/3) \crbig

{\cal V}_{(3, 6)} & = & {\cal V}_{(7/3, 14/3)} + V(3, 14/3) + V(2, 6) +
V(7/3,
6) 
+ V(3, 6) 

\end{array}
\eeq

In general, the potential at a given level has many extrema. Two of them
will 
be of particular interest for us:
\begin{enumerate}
\item We always find a translationally invariant minimum $\vec{T}_{vac}$
corresponding to the 
tachyon condensation. 
At this minimum, all fields with nonzero momentum have zero vev. We will 
use this solution when calculating the ratio $r^{(2)}$ defined in
eq.\refb{soption}.
\item If we start the numerical algorithm with initial values near 
$t_0 \approx 0.25$ and $t_1 \approx - 0.4$ then our numerical algorithm 
converges to the one-lump solution $\vec{T}_{lump}$ that we are interested
in.
\end{enumerate}

\begin{table}
\begin{center}\def\st{\vrule height 3ex width 0ex}
\begin{tabular}{|c|c|c|c|c|c|} \hline

Field & $(1/3, 2/3)$ & $(4/3, 8/3)$ & $(2, 4)$ &
$(7/3, 14/3)$ & $(3, 6)$ 
\st\\[1ex] \hline \hline

$t_0$ & 0.181034 & 0.214757 & 0.25703 & 0.265131 & 0.269224 \st\\[1ex]
\hline

$t_1$ & -0.344389 & -0.343566 & -0.384575 & -0.394396 & -0.394969
\st\\[1ex]
\hline

$t_2$ &... & -0.0955972 & -0.107424 & -0.12046 & -0.125011 \st\\[1ex]
\hline

$u_0$ &... &... & 0.0888087 & 0.0900609 & 0.0969175 \st\\[1ex] \hline

$v_0$ &... &... & -0.00675676 & -0.0175367 & -0.0172906 \st\\[1ex] \hline

$w_0$ &... &... & 0.0317837 & 0.0299617 & 0.0320394 \st\\[1ex] \hline

$u_1$ &... &... &... & -0.0643958 & -0.0648543 \st\\[1ex] \hline

$v_1$ &... &... &... & 0.0540447 & 0.0505836 \st\\[1ex] \hline

$w_1$ &... &... &... & -0.0187778 & -0.0189058 \st\\[1ex] \hline

$z_1$ &... &... &... & -0.0698363 & -0.0665402 \st\\[1ex] \hline 

$t_3$ &... &... &... &... & -0.0142169 \st\\[1ex] \hline

\end{tabular}
\end{center}
\caption{The values of various modes of the string field at the stationary 
point of the potential for $R=\sqrt{3}$ calculated at
various levels of approximation.}
\label{enable} 
\end{table}

\begin{table}
\begin{center}\def\st{\vrule height 3ex width 0ex}
\begin{tabular}{|l||l|l|} \hline
Level & $r^{(1)}$ & $r^{(2)}$ \st\\[1ex]\hline \hline
$(1/3; 2/3)$ & 1.32002 & 0.77377 \st\\[1ex] \hline
$(4/3; 8/3)$ & 1.25373 & 0.707471 \st\\[1ex] \hline
$(2; 4)$ &  1.11278 & 1.02368 \st\\[1ex] \hline
$(7/3; 14/3)$ & 1.07358 & 0.984467 \st\\[1ex] \hline
$(3, 6)$ & 1.06421 & 0.993855 \st\\[1ex] \hline
\end{tabular}
\end{center}
\caption{The ratio of the calculated mass of the
lump to the mass of the D0 brane in the two
schemes described in equations \refb{foption} and \refb{soption}.}
\label{ratiosc} 
\end{table} 

The solution $\vec T_{vac}$ can be found in refs.~\cite{9912249,0002237}.
In table~\ref{enable} we give the solutions $\vec{T}_{lump}$ at various
truncation 
levels. 
Having found $\vec{T}_{vac}$ and $\vec{T}_{lump}$ we can now calculate the
ratio of the lump 
mass to the D0-brane mass using the 
two different methods (\ref{foption}) and
(\ref{soption}). 
The results are given in Table~\ref{ratiosc}.
We see that the first method gives a monotonically decreasing lump mass 
whereas the second method is oscillating but gives a lump mass much 
closer to the expected mass. 

\begin{figure}[!ht] 
\leavevmode
\begin{center}
\epsfbox{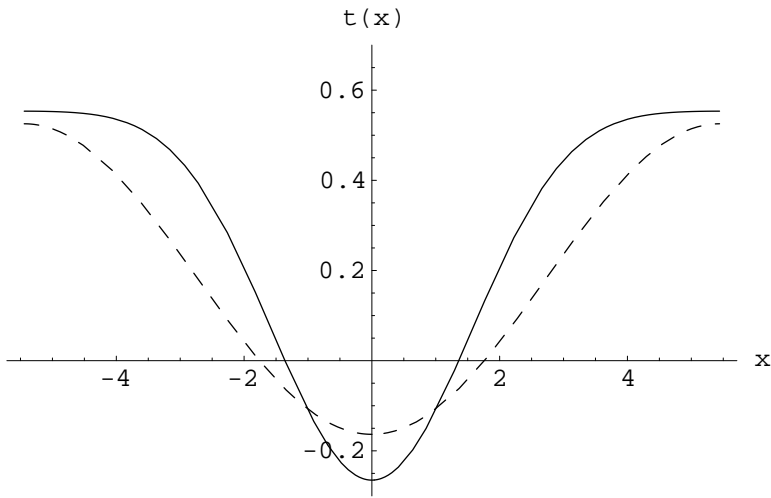}
\end{center}
\caption{The dashed line shows a plot of $t(x)$ 
for $R=\sqrt{3}$
at level (1/3, 2/3) approximation. The solid line  shows the plot of
$t(x)$ 
for $R=\sqrt{3}$ at the level (3,6)
approximation.} \label{f21}
\end{figure}

\begin{figure}[!ht] 
\leavevmode
\begin{center}
\epsfbox{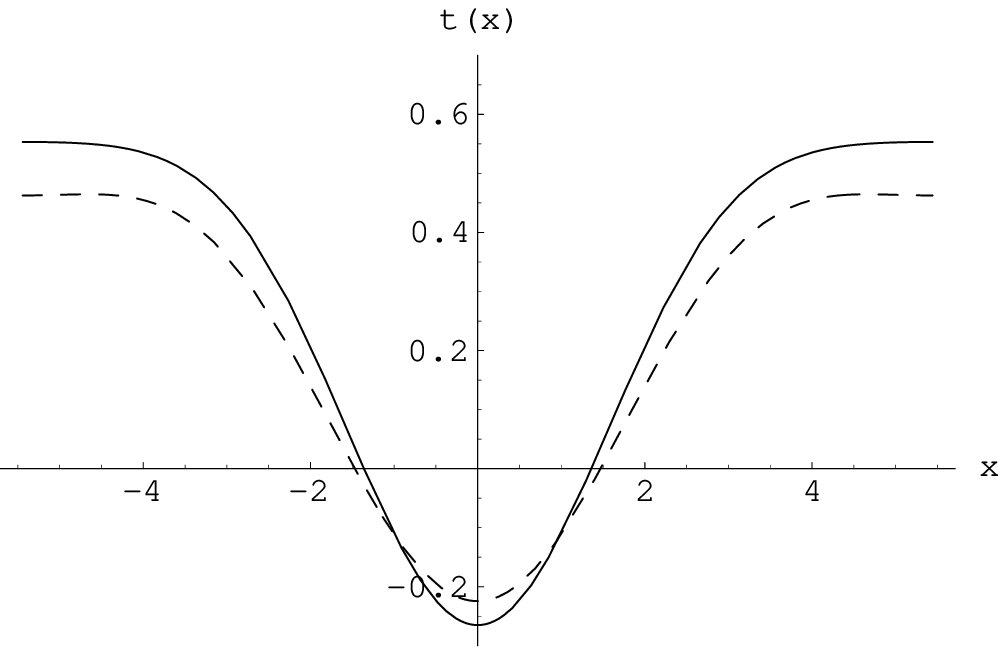}
\end{center}
\caption{The dashed line shows a plot of $t(x)$ 
for $R=\sqrt{3}$
at level (4/3, 8/3) approximation. The solid line  shows the plot of
$t(x)$ 
for $R=\sqrt{3}$ at the level (3,6)
approximation.} \label{f22}

\end{figure}
\begin{figure}[!ht] 
\leavevmode
\begin{center}
\epsfbox{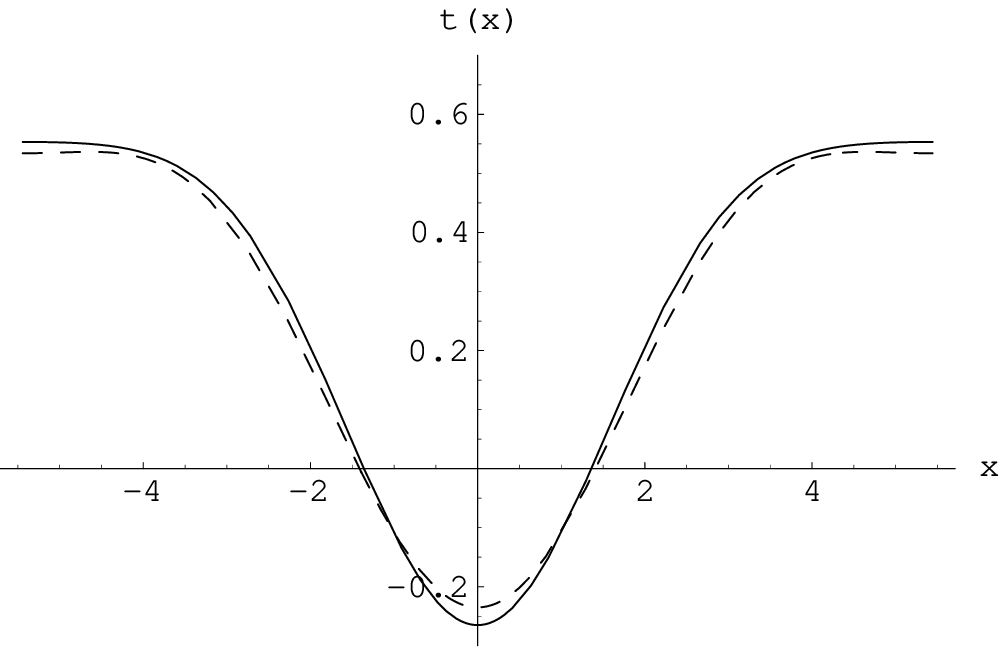}
\end{center}
\caption{The dashed line shows a plot of $t(x)$ 
for $R=\sqrt{3}$
at level (2, 4) approximation. The solid line  shows the plot of
$t(x)$ 
for $R=\sqrt{3}$ at the level (3,6)
approximation.} \label{f23}
\end{figure}

\begin{figure}[!ht] 
\leavevmode
\begin{center}
\epsfbox{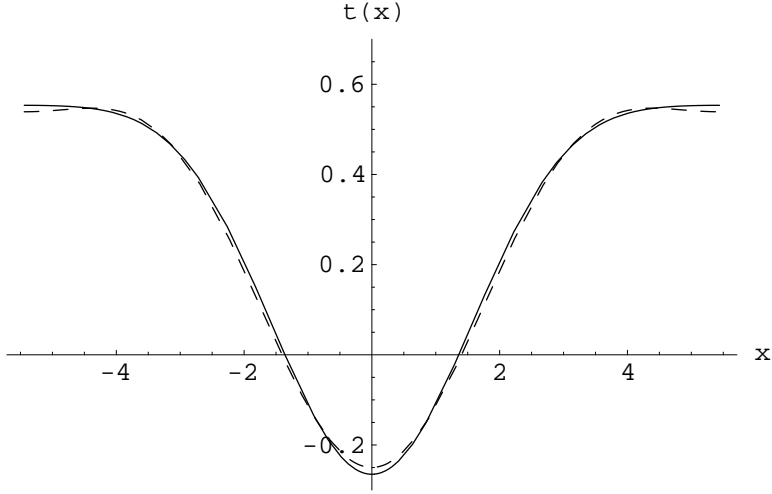}
\end{center}
\caption{The dashed line shows a plot of $t(x)$ 
for $R=\sqrt{3}$
at level (7/3, 14/3) approximation. The solid line  shows the plot of
$t(x)$ 
for $R=\sqrt{3}$ at the level (3,6)
approximation.} \label{f24}
\end{figure}

It is instructive to plot the profile of the tachyon field:
\be \label{etx}
t(x) = \sum_{n} t_n \cos{n x \over R} \, ,
\ee
as a function of $x$ and compare them at
different approximations. In figs.\ref{f21}-\ref{f24} we have plotted the
tachyon profiles at the level (1/3,2/3), (4/3,8/3), (2,4) and (7/3,14/3)
approximation respectively, each of them being 
superimposed on the tachyon profile at
the level (3,6) approximation.

For future use, we shall now define two new functions $F_0$ and $G_0$ as
follows:
\ben \label{ef0}
F_0(t_0, t_1, t_2, u_0, v_0, w_0, u_1, v_1, w_1, z_1;  R) 
&=& \VV_{({7\over 3},
{14\over 3})}\, , \nonumber \\
G_0(t_0, t_1, t_2, u_0, v_0, w_0, u_1, v_1, w_1, z_1;  R) 
&=& \VV_{(3,6)} - V(3,6) - V(2, 14/3) - V(7/3, 4) \nonumber \\
&& - V(7/3, 6) -
V(3, 14/3)
\een
where $\VV_{({7\over 3}, {14\over 3})}$ and $\VV_{(3,6)}$ have been
defined in
eqs.\refb{v0}-\refb{vtot}.
The right hand side of this equation has to be interpreted as
a
function of the various modes $t_0,\ldots z_1$ and $R$, {\it without $R$
being set to $\sqrt 3$.} The function $F_0$ and $G_0$ defined here will be
useful in
constructing the potential $\VV_{(M,N)}$ for other values of $R$, as will
be
discussed in the next section.

\sectiono{Tachyon Lump at Other Radii} \label{s3}

In this section we shall discuss the construction of the tachyonic lump
solution on circles of radii other than 
$\sqrt{3}$, and compare the results
with those obtained for $R=\sqrt{3}$. As the basic techniques have already
been discussed in the previous two sections, in this section we shall only
quote the results. 

\subsection{$R>\sqrt{3}$}

First we need to decide which values of 
$R$ we shall use to study the lump.
Although this choice is arbitrary, 
there is slight simplification of counting
levels if we choose $R$ such that the level 
of $u_1$, $v_1$, $w_1$ and $z_1$
coincide with that of one of the 
harmonics (say $t_n$) of the tachyon field.
This requires
\be \label{er1}
2 + {1\over R^2} = {n^2 \over R^2}, 
\qquad \to \qquad R = \sqrt{n^2 -1\over
2}\, .
\ee
We shall consider the values $n=4,5,6$ 
corresponding to $R=\sqrt{15\over 2},
\sqrt{12}, \sqrt{35\over 2}$. In each case we shall be using the level
$(2+{1\over R^2}, 4+{2\over R^2})$ approximation to the potential. For
this
we need to include up to the $n$-th harmonic 
of the tachyon field $t$ and the
first harmonics of the fields $u,v,w$ and $z$.

For these additional $R$ values all interactions 
present in $\VV_{({7\over 3}, {14\over 3})}$ at
$R= \sqrt{3}$ are still present. We need, however,
further interactions as can be checked using the
generating function \refb{genfunc}. These additional
interactions can be expressed in terms of the
following functions:
\ben \label{ef1}
&& F_1(t_0, \cdots,  t_4, u_0, v_0, w_0, u_1, v_1, w_1, z_1;  R) 
\nonumber \crbig
&=&   
- {1 \over 4} \bigg(1 - {9\over R^2}\bigg) t_3^2 
- {1\over 4} \bigg(1 - {16 \over R^2}\bigg) t_4^2 
\nonumber \\
&&\hskip-22pt
+ {1 \over 2} K^{3 - 18/ R^2} t_0 t_3^2 
+ {1 \over 2} K^{3 - 32/ R^2} t_0 t_4^2 
+ {1 \over 2}  K^{3 - 14/R^2} t_1 t_2 t_3
+ {1\over 4} K^{3 - 24/R^2} t_2^2 t_4
+ {1 \over 2} K^{3 - 26/R^2} t_1 t_3 t_4 
 \nonumber \\ &&
\hskip-22pt+ \bigg( {11 \over 32}u_1 
 -{125\over 64} w_1
+ \Bigl( {25\over 2 R^4} - {11 \over 16 R^2} \Bigr)
 z_1
+ \Bigl({11 \over 4 R^2} - {5 \over 64}\Bigr) v_1\bigg)  K^{1 - 14/R^2}
t_2 t_3
\nonumber\\
\een
\ben \label{ef2}
&& F_2(t_0, \cdots,  t_5, u_0, v_0, w_0, u_1, v_1, w_1, z_1;
R)
\nonumber
\crbig &=& - {1\over 4} \bigg(1 - {25\over R^2}\bigg) t_5^2 +
{1\over 2} K^{3 - 38/R^2}\,t_2\,t_3\,t_5 + {1 \over 2} 
K^{3 - 42/R^2}\,t_1\,t_4\,t_5 + 
{1\over 2} K^{3 - 50/R^2}\,t_0\,t_5^2 \nonumber \\
&& + \bigg(\, {11 \over 32 }u_0 +  \Bigl(-{5 \over 64} + {9 \over 2
R^2}\Bigr)v_0 
- {125\over 64}  w_0  \bigg)  K^{1 -
18/R^2} 
 t_3^2 \nonumber \\
&& + \bigg( {11 \over 32}u_1  
+ \Bigl(-{5\over 64} + {23\over 4 R^2}\Bigr)\,v_1
 - 
{125\over 64} \,w_1
+ \Bigl( {49\over 2 R^4}  - 
{11\over 16 R^2} \Bigr)\, z_1 \bigg) \,K^{1 - 26/R^2}\,t_3\,t_4 \nonumber
\\
\een
\ben \label{ef3}
&& F_3(t_0, \cdots, t_6, u_0, v_0, w_0, u_1, v_1, w_1,
z_1;
R)
\nonumber
\crbig 
 &=& -{1\over 4} \bigg(1 - {36\over R^2}\bigg) t_6^2
+ {1\over 4}\,K^{3 - 54/R^2}\,t_3^2\,t_6 + {1\over 2}\,K^{3 -
56/R^2}\,t_2\,t_4\,t_6 + {1\over 2} 
K^{3 - 62/R^2}\,t_1\,t_5\,t_6 \nonumber \\
&& + {1\over 2}
K^{3 - 72/R^2}\,t_0\,t_6^2  + \bigg( {11\over 32}\, u_0 
 + \Bigl(-{5\over 64} +{8\over 
R^2}\bigg) \, v_0
-  {125\over 64}\, w_0 \bigg)  \,K^{1 - 32/R^2}\,t_4^2\, . \nonumber \\
\een

We shall now write down our results for level $(2+{1\over R^2}, 4+{2\over
R^2})$ approximation for the potential for $R^2=(n^2-1)/2$ in terms of the
functions $F_0,\ldots F_3$ defined in eqs.\refb{ef0},
\refb{ef1}-\refb{ef3}. These are as follows:
\ben \label{ep152}
&& \VV_{(32/15,64/15)}(t_0, \cdots, t_4, u_0, v_0, w_0, u_1, v_1,
w_1,
z_1;
R=\sqrt{15/2}) \nonumber \\ 
&=&  
F_0(t_0, t_1, t_2, u_0, v_0, w_0, u_1, v_1, w_1, z_1;
R=\sqrt{15/2}) \nonumber \\
&& + F_1(t_0,\cdots, t_4, u_0, v_0, w_0, u_1, v_1, w_1, z_1;
R=\sqrt{15/2})\, ,
\een
\ben \label{ep12}
&& \VV_{(25/12,25/6)}(t_0,\cdots, t_5, u_0, v_0, w_0, u_1,
v_1,
w_1,
z_1;
R=\sqrt{12}) \nonumber \\ 
&=&  
F_0(t_0, t_1, t_2, u_0, v_0, w_0, u_1, v_1, w_1, z_1;
R=\sqrt{12}) \nonumber \\
&& + F_1(t_0, \cdots, t_4, u_0, v_0, w_0, u_1, v_1, w_1, z_1;
R=\sqrt{12})\nonumber \\
&& + F_2(t_0, \cdots, t_5, u_0, v_0, w_0, u_1, v_1, w_1, z_1;
R=\sqrt{12})\, ,
\een
\ben \label{ep352}
&& \VV_{(72/35,144/35)}(t_0, \cdots, t_6, u_0, v_0, w_0,
u_1,
v_1,
w_1,
z_1;
R=\sqrt{35/2}) \nonumber \\ 
&=&  
F_0(t_0, t_1, t_2, u_0, v_0, w_0, u_1, v_1, w_1, z_1;
R=\sqrt{35/2}) \nonumber \\
&& + F_1(t_0,\cdots, t_4, u_0, v_0, w_0, u_1, v_1, w_1, z_1;
R=\sqrt{35/2})\nonumber \\
&& + F_2(t_0, \cdots, t_5, u_0, v_0, w_0, u_1, v_1, w_1, z_1;
R=\sqrt{35/2}) \nonumber \\
&& + F_3(t_0, \cdots, t_6, u_0, v_0, w_0, u_1, v_1, w_1,
z_1;
R=\sqrt{35/2})\, .
\een

As in the previous section, we can find a tachyonic lump solution by
starting with a non-zero seed value of $t_1$. 
The numerical solutions are given in Table~\ref{difrad}. 
The result for the two ratios $r^{(1)}$ and $r^{(2)}$, defined in
eqs.\refb{foption} and \refb{soption} are given in Table~\ref{ratiodr}.

\begin{table}
\begin{center}\def\st{\vrule height 3ex width 0ex}
\begin{tabular}{|c|c|c|c|c|} \hline

Field & $R=\sqrt{15/2}$ & $R=\sqrt{12}$ & $R=\sqrt{35/2}$ &
$R=\sqrt{11/10}$ 
\st\\[1ex] \hline \hline

$t_0$ & 0.363333 & 0.401189 & 0.424556 & 0.0804185 \st\\[1ex]
\hline

$t_1$ & -0.308419 & -0.255373 & -0.218344 & -0.31707 \st\\[1ex]
\hline

$t_2$ & -0.19463 & -0.190921 & -0.176679  & -0.00983574 \st\\[1ex]
\hline

$t_3$ &-0.0849552 & -0.122721 & -0.132269 & ... \st\\[1ex] \hline

$t_4$ & -0.0248729 & -0.0575418 & -0.0830114 & ... \st\\[1ex] \hline

$t_5$ &... & -0.0210929 & -0.0409281 & ... \st\\[1ex] \hline

$t_6$ &... &... & -0.0178687 & ... \st\\[1ex] \hline

$u_0$ & 0.118792 & 0.131499 & 0.139048 & 0.0318155 \st\\[1ex] \hline

$v_0$ & 0.0131977 & 0.020668 & 0.0263317 & -0.0591248 \st\\[1ex] \hline

$w_0$ & 0.0380389 & 0.0417417 & 0.0438076 & 0.0132021 \st\\[1ex] \hline

$u_1$ & -0.0712708 & -0.0629211 & -0.0567058 & -0.0052739 \st\\[1ex]
\hline

$v_1$ & -0.0958004 & -0.0657449 & -0.0476215 & -0.0119114\st\\[1ex] \hline

$w_1$ & -0.0181708 & -0.0150031 & -0.0131234 & -0.000863176 \st\\[1ex]
\hline

$z_1$ & 0.0860302 & 0.058747 & 0.0418645 & 0.00570249 \st\\[1ex] \hline 

\end{tabular}
\end{center}
\caption{The values of various modes of the string field at the stationary 
point of the potential for different radii.}
\label{difrad} 
\end{table}

\begin{table}
\begin{center}\def\st{\vrule height 3ex width 0ex}
\begin{tabular}{|l||l|l|} \hline
R & $r^{(1)}$ & $r^{(2)}$ \st\\[1ex]\hline \hline
$\sqrt{15/2}$ & 1.14625 & 1.00535 \st\\[1ex] \hline
$\sqrt{12}$ & 1.19147 & 1.01324 \st\\[1ex] \hline
$\sqrt{35/2}$ &  1.23876 & 1.02353 \st\\[1ex] \hline
$\sqrt{11/10}$ &  1.02175 & 0.979149 \st\\[1ex] \hline
\end{tabular}
\end{center}
\caption{The ratio of the calculated mass of the
lump to the mass of the D0 brane at various
radii in the two
schemes  described in equations \refb{foption} and \refb{soption}.}
\label{ratiodr} 
\end{table}

\begin{figure}[!ht] 
\leavevmode
\begin{center}
\epsfbox{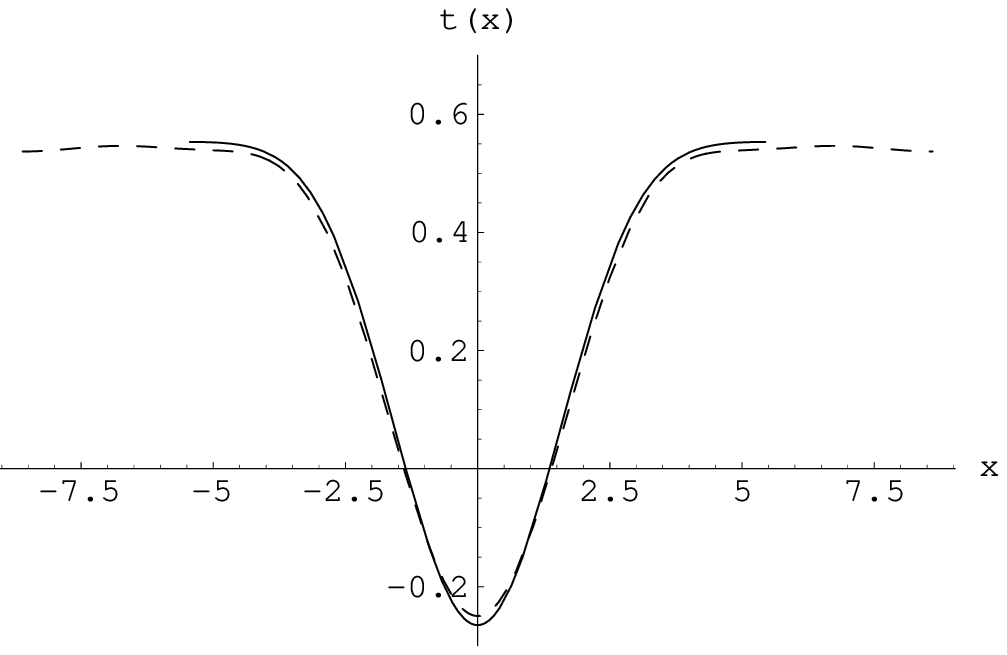}
\end{center}
\caption{The dashed line shows a plot of $t(x)$ 
for $R=\sqrt{15/2}$
at level (32/15, 64/15) approximation. The solid line spanning a
smaller range of $x$ shows the plot of $t(x)$ 
for the level (3,6)
approximation at $R=\sqrt{3}$.} \label{f31}
\end{figure}

\begin{figure}[!ht] 
\leavevmode
\begin{center}
\epsfbox{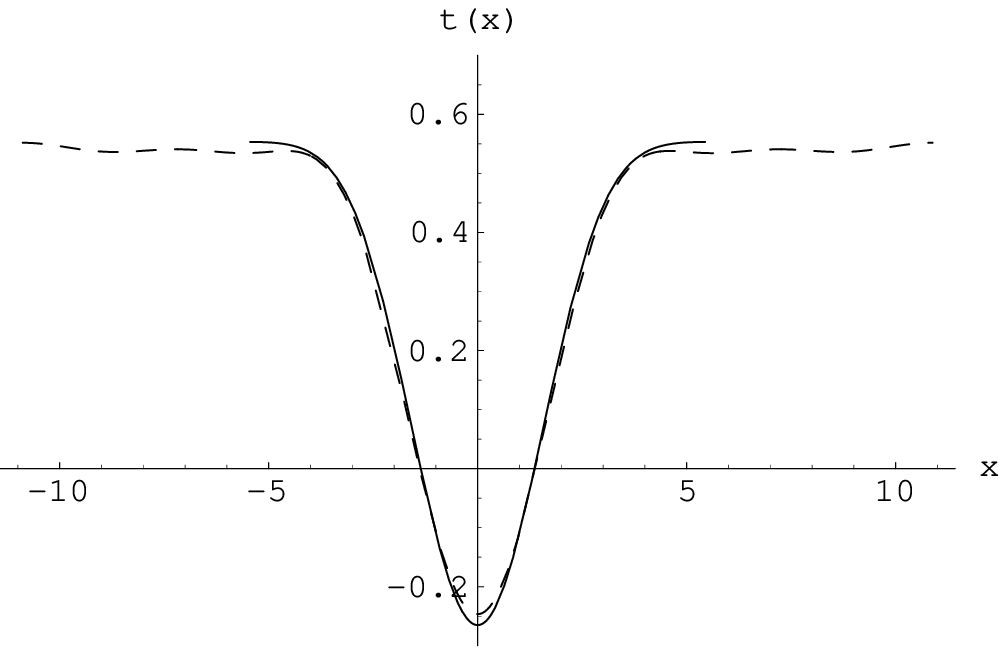}
\end{center}
\caption{The dashed line shows a plot of $t(x)$ 
for $R=\sqrt{12}$
at level (25/12, 25/6) approximation. The solid line spanning a
smaller range of $x$ shows the plot of $t(x)$ 
for the level (3,6)
approximation at $R=\sqrt{3}$.} \label{f32}
\end{figure}

\begin{figure}[!ht] 
\leavevmode
\begin{center}
\epsfbox{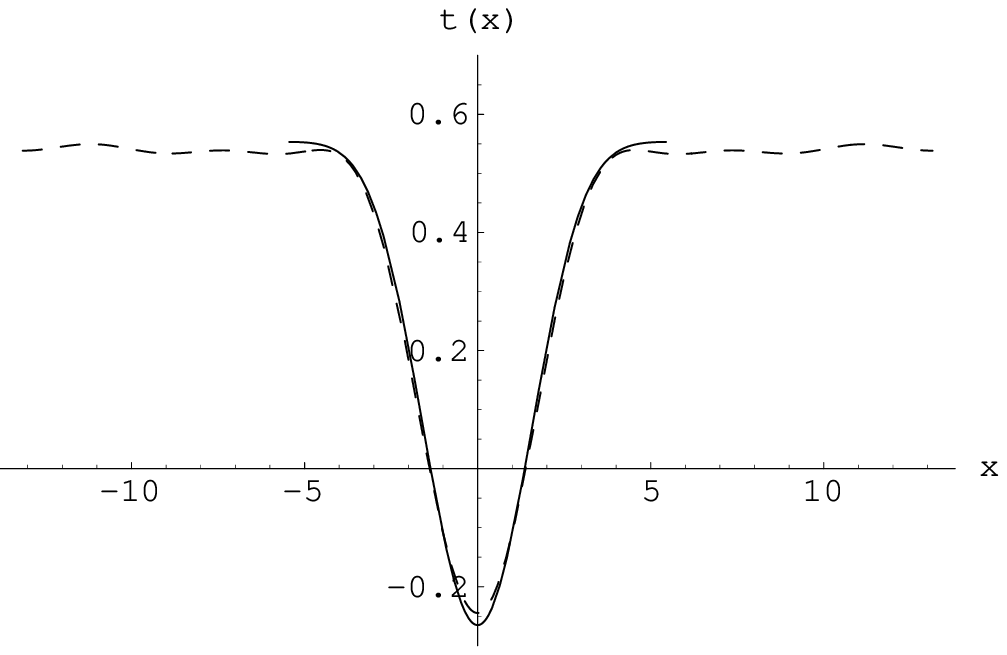}
\end{center}
\caption{The dashed line shows a plot of $t(x)$ 
for $R=\sqrt{35/2}$
at level (72/35, 144/35) approximation. The solid line spanning a
smaller range of $x$ shows the plot of $t(x)$ 
for the level (3,6)
approximation at $R=\sqrt{3}$.} \label{f33}
\end{figure}

In Figs.\ref{f31}-\ref{f33} we have plotted the tachyon field $t(x)$
defined in eq.\refb{etx}
as a
function of $x$ for each of the three values of $R$. For reference we have
also plotted on the same graph the function $t(x)$ obtained in the level
(3,6) approximation for $R=\sqrt{3}$. As is seen from these figures, the
tachyon profiles for different radii are almost undistinguishable from
each other even though they are obtained as superpositions of harmonics of
very different wave-lengths.

\subsection{$R<\sqrt{3}$}

Finally we would like to study how the shape of the soliton changes when
$R$ is small. For this we take $R=\sqrt{1.1}$ and work at the level
(40/11,
80/11) approximation of the potential. One can show that to this level of
approximation the potential is given by,
\ben \label{ersmall}
&& \VV_{(40/11, 80/11)}(t_0, t_1, t_2, u_0, v_0, w_0, u_1, v_1, w_1, z_1;
R=\sqrt{1.1})\nonumber \\ 
&=& G_0(t_0, t_1, t_2, u_0, v_0, w_0, u_1, v_1, w_1, z_1;  R=\sqrt{1.1})\,
,
\een
where $G_0$ has been defined in eq.\refb{ef0}. The tachyonic lump solution
for this potential is given in table~\ref{difrad}.
The results for the two ratios $r^{(1)}$ and $r^{(2)}$ defined in
eqs.\refb{foption} and \refb{soption} are given in table~\ref{ratiodr}.

\begin{figure}[!ht] 
\leavevmode
\begin{center}
\epsfbox{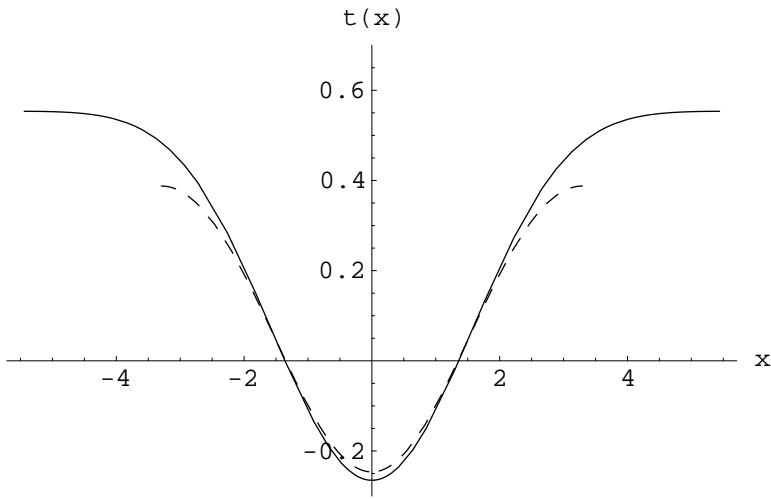}
\end{center}
\caption{The dashed line shows a plot of $t(x)$ 
for $R=\sqrt{11/10}$
at level (40/11, 80/11) approximation. The solid line spanning a
smaller range of $x$ shows the plot of $t(x)$ 
for the level (3,6)
approximation at $R=\sqrt{3}$.} \label{f35}
\end{figure}

We have displayed in fig.~\ref{f35} the tachyon profile, superimposed on
the tachyon profile for the level (3,6) approximation at $R=\sqrt{3}$. As
can be seen from this figure, for $R=\sqrt{11/10}$ there is not enough
room for the full lump solution to fit in, but the profile of the lump
at smaller radius follows closely the profile at larger radius near the
core.

\subsection{Size of the lump}

We can
estimate the size of the lump at different radii in a 
somewhat systematic way by fitting the lump profile with 
a gaussian curve of the form:
\be \label{gausscurve}
G(x) = a + b \cdot e^{-x^2 / (2 \sigma^2)}\, .
\ee
We calculate the parameters $a$, $b$ and $\sigma$ using a nonlinear
regression 
algorithm on a set of points chosen on the lump profile in the following 
way. For $R\ge \sqrt{3}$:

\begin{itemize}
\item  We take 100 points, regularly spaced in $x$, in
the core of the lump 
from $x = - \sqrt{3}\, \pi$ to $x =  \sqrt{3}\, \pi$.

\item We take a smaller density of points, regularly spaced in $x$, on 
the rest 
of the circle (where the profile is essentially flat). Here we have taken 
20, 30 and 40 points for $R = \sqrt{15/2}$, $R = \sqrt{12}$ and 
$R = \sqrt{35/2}$ respectively. 

\end{itemize}

\noindent In the case of $R = \sqrt{11/10}$, we take 100 points from 
$x = - \sqrt{11/10} \, \pi$
to  $x =  \sqrt{11/10} \, \pi$

\begin{table}
\begin{center}\def\st{\vrule height 3ex width 0ex}
\begin{tabular}{|c|c|c|c|} \hline
Radius & a & b & $\sigma$  \st\\[1ex]\hline \hline
$\sqrt{3}$ & 0.559814 & -0.828599 & 1.52341 \st\\[1ex] \hline
$\sqrt{15/2}$ & 0.546313 & -0.804112 & 1.5595 \st\\[1ex] \hline
$\sqrt{12}$ & 0.544226 & -0.801652 & 1.54089 \st\\[1ex] \hline
$\sqrt{35/2}$ & 0.54328 & -0.799957 & 1.54477 \st\\[1ex] \hline
$\sqrt{11/10}$ & 0.451678 & -0.702596 & 1.41847 \st\\[1ex] \hline 
\end{tabular}
\end{center}
\caption{The result for the best fit of the profile of the lump with the
gaussian curve described in eq.\refb{gausscurve}.} \label{regress}
\end{table}

The results of the regression at the different radii are given in
table~\ref{regress}.
We see that the size of the lump, which can be defined as a multiple of 
$\sigma$, is essentially independent of the radius (it increases by about 
1.5 \% when R increases from $\sqrt{3}$ to $\sqrt{35/2}$). Even when 
there is not enough room for the lump to fit in ($R = \sqrt{11/10}$), 
the lump is 
only slightly compressed (by about 7 \%). A reasonable 
definition for the size would be $6 \, \sigma$, with the solution
extending by $3 \, \sigma$ both along the positive and the negative
$x$-axis.
With this convention, the 
lump will have a size of approximately $9.3\, \sqrt{\alpha^{\prime}}$.
This is close to the answer obtained in ref.\cite{0002117}.

\sectiono{Conclusions and Open Questions} \label{s4}

In this paper we have developed and tested the level
expansion method in string field theory beyond translationally
invariant vacuum solutions.  This enabled us to give
a systematic method for calculating quantities related to
tachyon lumps and to give an accurate description of
D-branes as  tachyonic lumps in bosonic string field theory.
Given the accuracy of our calculations (about 1\% typically)
we are confident that the profile of the lump that we have found is 
indeed very close to the exact one. As we have seen, 
as long as 
the radius is sufficiently big the lump
has a definite radius independent profile. Indeed,
when approximated by a gaussian,
the lump representing a D-brane has $\sigma \simeq 1.55\sqrt{\alpha'}$.  We
also
considered the profile of the
tachyon  lump for $R= \sqrt{1.1\alpha'}$, a radius
sufficiently small that the large radius profile of the lump does not fit
on the circle. We saw that 
the bottom part of the lump is essentially 
unchanged.

There are some questions related to the
present work that we have not
addressed. In particular we have not
produced a lump solution in string field theory for $R=1$, where the 
tachyon harmonic $t_1$ becomes exactly marginal
and the D0 and D1 branes have the same mass.
Presumably, for small $(R^2-1)$ one must go
fairly high in the level expansion to produce
an accurate description. We have also not discussed
the case $R<1$, where the D0 brane is unstable against
decay into the D1 brane, or into the 
translationally invariant vacuum.  We have also not
tried to describe several D0 branes, all located
at the same position. 

We have not  discussed issues related to
the size of the lump representing a D-brane.
While in the conformal field theory description a D-brane is an object
with
a well defined position, in string field theory it is a 
fat object, with thickness of the order of the string scale. Since string
field theory
is a gauge theory one may wonder if the size
is an artifact of the chosen gauge.  We do not at present know the answer
to this question.
The simplest way to get some insight into the nature of
this extended solution would be to try to find
out the energy density. This fails since the string field theory action is 
nonlocal, and hence
there is no known expression for energy density in this theory.
It would be interesting to examine some physical
question that could help interpret the nature 
of this size~\cite{9608024}. According to the conjectures of
refs.\cite{9902105,RECK}, all physical quantities calculated in the
background of the lump solution must agree with those calculated in the
background of a lower dimensional D-brane.

The methods used in this paper should be able 
to deal with:

\begin{itemize}

\item Neveu-Schwarz string field 
theory, where tachyon kinks rather than lumps represent
lower dimensional D-branes. One way
to deal with the boundary conditions on a circle would
be to place both a kink and an anti-kink at diametrically opposite points 
of the circle. Another, probably
more efficient way would
be to include a Wilson line along the circle in such a 
way that the tachyon boundary conditions are twisted~\cite{9808141}.

\item  Higher codimension D-branes.  In~\cite{0002117} it was observed
that as the codimension is increased 
the naive use of the tachyon ``bounce" gave increasingly
worse approximations to the lump mass. We believe that our
methods will enable calculations to any desired accuracy.
The simplest situation would involve making two of the original
brane dimensions into circles and including harmonics in both
directions by simple extension of the methods of section  2.2. 

\item Intersecting D-branes. The simplest setup would be to 
begin with a D2-brane on a torus and generate a pair of 
transverse D1 branes intersecting at one point.

\end{itemize} 

We hope that our analysis will ultimately provide a 
more refined understanding
of string field theory and its geometry. One application 
is already apparent; if we
could get a formulation of string field theory around the 
translationally invariant vacuum where the original 
D-brane is no longer present, such formulation
will have more unbroken symmetries than the current
formulation.

It is interesting to note that the level expansion
method used here incorporates into the calculational
scheme an ultra-violet (UV) cutoff. Since $l = p^2 + \cdots$, working
at fixed $l$ implies a upper bound to the momentum (in the spatial
directions). {}From this one is naturally led to
propose a level expansion method for {\it quantum string field
theory}. One approach could be to use the Euclidean version of the
theory, and make periodic all 
directions including time\cite{9305139}, thus turning, at any fixed level
$M$, the set
of all relevant fields into a set of expansion coefficients $c_n$, with 
$l(|\phi_n\rangle)\le M$.  Since we are setting the whole system
in a box, we also have a natural infra-red cutoff. 
The whole quantum path integral 
$\int \prod [dc_n] \exp (-S(c_n)/\hbar)$ could
then be evaluated.\footnote{Here $S$ should be the truncation, to the
given level of approximation, of the
full
quantum action satisfying the exact quantum Batalin-Vilkovisky master
equation. It is not clear that the cubic open string field theory
provides such solution\cite{thorn}. The open-closed string field theory
introduced in ref.~\cite{9705241} does provide a well defined quantum
action, but due to the non-polynomial nature of the action it is not clear
how to carry out the level expansion in this
theory.} Alternatively,
one could make all dimensions except time periodic. In this case
the result would be the quantum mechanics of the wave functions $c_n(t)$.
It would be exciting if the level expansion gave
a concrete calculational definition of quantum string field theory,
a definition one could in practice feed to a computer in order to 
calculate observables to any desired degree of precision.   

\bigskip
\noindent {\bf Acknowledgements}: We would like to thank R.~Gopakumar,
S.~Minwalla, L.~Rastelli and W.~Taylor for
discussions. A.S. acknowledges the Physics department of  Harvard
University and the Center for Theoretical Physics at MIT for hospitality
during part of this work. 
The work of  N.M. and B.Z. was supported in part
by DOE contract \#DE-FC02-94ER40818.

\end{document}